\documentclass[12pt,epsf]{article}
\usepackage{enumerate,amsmath,amscd,amssymb,graphicx
}

\makeatletter
\@addtoreset{equation}{section}

\renewcommand{\thefootnote}{\fnsymbol{footnote}}
\makeatother

\def\bm#1{\mbox{\boldmath $#1$}}
\newcommand{\Df}[1]{\overset{(4)}{#1}}
\newcommand{\Ds}[1]{\overset{(5)}{#1}}

\begin{document}

\thispagestyle{empty}
%
\begin{flushright}
TIT/HEP--518 \\
hep-th/0404114 \\
April, 2004 \\
\end{flushright}
\vspace{3mm}
\begin{center} 
{\bf 
Non-BPS Walls and Their Stability in 5D Supersymmetric Theory
}
\vspace{5mm}

  {\bf 
  Minoru~Eto~$\!\!^{a}\!\!$}
\footnote{\it  e-mail address: 
meto@th.phys.titech.ac.jp
}~
  {\bf 
  Nobuhito~Maru~$\!\!^{b}\!\!$}
\footnote{\it  e-mail address: 
maru@postman.riken.go.jp, Special Postdoctoral Researcher
}  
~and~~ {\bf 
Norisuke~Sakai~$\!\!^{a}\!\!$}
\footnote{\it  e-mail address: 
nsakai@th.phys.titech.ac.jp
}

\vskip 1.5em

{ \it 
  $^{a}$Department of Physics, Tokyo Institute of 
Technology \\
Tokyo 152-8551, JAPAN  \\
 $^{b}$Theoretical Physics Laboratory, RIKEN \\ 
 Saitama 351-0198, JAPAN 
   }
\vspace{20mm}
{\bf Abstract}\\[5mm]
{\parbox{13cm}{\hspace{5mm}
An exact solution of non-BPS multi-walls 
is found in supersymmetric massive 
$T^\star(\mathbb{CP}^1)$ model 
in five dimensions. 
The non-BPS multi-wall solution is found to have 
no tachyon. 
Although it is only metastable under large fluctuations, 
we can give topological stability by considering 
a model with a double covering of the 
$T^\star(\mathbb{CP}^1)$ target manifold. 
The ${\cal N}=1$ supersymmetry preserved 
on the four-dimensional world volume of one wall 
is broken by the coexistence of the other wall. 
The supersymmetry breaking is exponentially 
suppressed as the distance between the walls increases. 
}}
\end{center}
\vfill
\newpage
\setcounter{page}{1}
\setcounter{footnote}{0}
\renewcommand{\thefootnote}{\arabic{footnote}}

\section{Introduction}

Brane-world scenario with 
extra dimensions\cite{HoravaWitten}, \cite{LED}, 
\cite{RandallSundrum} have attracted much attention 
in recent years. 
It has also been useful to implement supersymmetry 
(SUSY) to obtain solitons such as walls 
where particles should be localized. 
SUSY has been most useful to obtain 
realistic unified theories\cite{DGSW}, but 
one of their least understood problems 
is the origin of the SUSY breaking. 
Half of SUSY 
can be preserved by 
walls\cite{cvetic1}.  
Then they automatically 
become minimal energy solutions with given 
boundary conditions\cite{Bogomolny:1975de}, 
\cite{WittenOlive}. 
They are called the Bogomolo'nyi-Prasad-Sommerfield (BPS) 
states and 
are assured of stability by the central charge 
of the SUSY algebra\cite{WittenOlive}. 
It has been found that the coexistence of these walls 
can break SUSY completely, leading to a possible 
origin of the SUSY breaking\cite{Maru:2000sx}. 
On the other hand, the stability of 
such non-BPS configurations 
are no longer guaranteed by SUSY. 
If we introduce topological quantum numbers 
such as the winding number by taking appropriate 
target manifold and by compactifying 
the extra dimension, we can have a 
stable non-BPS walls\cite{Maru:2001gf}, \cite{SS}. 
By considering the SUSY sine-Gordon model 
with $\pi_1(S^1)={\bf Z}$, stable non-BPS 
multi-wall configurations 
have been obtained in four-dimensional theories
\cite{Maru:2001gf}. 
However, models in five dimensions are needed 
to obtain realistic models for brane-world. 
In five dimensions or higher, we need to implement 
at least eight supercharges. 
Theories with eight SUSY are quite restricted. 
The simplest multiplet in such theories is called 
hypermultiplet which contains spin $0$ and $1/2$ only. 
In order to obtain interacting hypermultiplets, we need to 
have nonlinear kinetic terms or gauge interactions
\cite{Alvarez-Gaume:1983ab}, \cite{SierraTownsend}. 
The nonlinear sigma model should have a hyper-K\"ahler 
target manifold~\cite{RT}--\cite{Gibbons:1987pk} 
with appropriate potential, 
which are called massive hyper-K\"ahler 
nonlinear sigma models. 
The BPS wall solutions of hypermultiplets 
have been obtained for the simplest of 
such nonlinear sigma models, a massive 
$T^\star(\mathbb{CP}^1)$ model\cite{Arai:2002xa},
\cite{Arai:2003es}. 
Even the BPS $n$-wall 
solutions have been constructed 
in $T^\star(\mathbb{CP}^n)$ for $n \ge 1$ 
\cite{GTT}. 
They have been successfully embedded into supergravity 
in five dimensions\cite{AFNS}, \cite{EFNS}. 
However, no non-BPS multi-wall solutions have been 
obtained so far.

The purpose of our paper is to present 
non-BPS multi-wall solutions in a SUSY theory in five 
dimensions and to discuss their stability. 
We show that the $T^\star(\mathbb{CP}^1)$ nonlinear sigma model 
in five dimensions admits exact solutions of 
non-BPS multi-walls which are 
identical to those found in four dimensions. 
We find that the small fluctuations around 
these solutions have no tachyons. 
This result implies that the non-BPS walls are stable 
under small fluctuations. 
However, the target manifold of the nonlinear sigma 
model, $T^\star(\mathbb{CP}^1)$, does not admit 
winding number 
as a topological quantum number, since the homotopy 
group is trivial ($\pi_1(T^\star(\mathbb{CP}^1))=0$), 
contrary to the target manifold 
of the sine-Gordon nonlinear sigma model in four dimensions. 
By using a variational approach, we demonstrate 
that the non-BPS multi-wall configurations 
can be continuously deformed into an energetically lower 
configuration with no walls. 
In conformity with the local stability of the background, 
the energy of the configuration achieves a local minimum 
at the non-BPS multi-wall solution and exhibits a maximum 
before reaching the no wall configuration 
at large deformations. 
Therefore the non-BPS multi-wall solutions are only 
metastable in the $T^\star(\mathbb{CP}^1)$ model. 
Although the metastability may be sufficient 
for the brane-world to exist during the finite 
cosmological lifetime
, we can give 
a topological stability to the non-BPS multi-wall 
solutions by considering a model 
with the target space of a double cover of 
$T^\star(\mathbb{CP}^1)$, which admits a nontrivial homotopy group 
$\pi_1(S^1)={\bf Z}$. 
Since the local properties of the double cover is 
identical to those of $T^\star(\mathbb{CP}^1)$, it should 
satisfy all the requirements of hyper-K\"ahler 
manifold needed for eight SUSY \cite{Alvarez-Gaume:1983ab}. 
No singularities or 
obstructions seem to occur even globally 
for the double cover. 
Small fluctuations around a 
background should not depend on global properties 
such as a generalization to the double cover. 
We believe that this may precisely be the reason 
for the fact that there is no tachyon around the 
non-BPS multi-wall background, since 
the small fluctuation analysis for 
$T^\star(\mathbb{CP}^1)$ should be 
identical to the case of 
double cover where the topological stability 
is guaranteed by the nontrivial homotopy 
group $\pi_1$. 

We also obtain spectra of massless bosons or light bosons 
which become massless in the limit of infinitely 
separated walls. 
They correctly form complex scalar fields 
needed to realize 
the chiral scalar multiplets of four SUSY. 
The SUSY breaking due to the coexistence of 
walls provides a mass difference between bosons 
and fermions in the chiral scalar multiplets. 
This is explicitly demonstrated for the supermultiplet 
with a massless fermion and a slightly massive boson, 
similarly to the case of 
four-dimensional models\cite{Maru:2000sx}, \cite{Maru:2001gf}. 
The mass splitting is found to decrease 
exponentially as the distance between walls 
increases. 
We can also embed our model and the solution into 
the five-dimensional supergravity. 
Similarly to the sine-Gordon model\cite{Eto:2003xq} 
and other models\cite{Eto:2003bn} in four dimensions, 
we expect that the coupling to gravity\cite{cvetic} does not 
introduce new instability, all the massless fields 
are absorbed by gauge fields via Higgs 
mechanism, and the lightest scalar field (radion) 
is nothing but the lightest massive mode of our solution in 
the rigid SUSY model, at least at weak gravitational 
coupling. 

In sect.\ref{sc:nonBPSsolution}, 
we describe the non-BPS multi-wall 
solutions both for the SUSY sine-Gordon model 
in four dimensions and for the $T^\star(\mathbb{CP}^1)$ 
model in five dimensions. 
In sect.\ref{sc:stability}, 
it is shown that the multi-wall solutions of 
 the $T^\star(\mathbb{CP}^1)$ 
model in five dimensions are stable 
under small fluctuations, but are deformable 
continuously to no wall configurations, 
showing their metastability. 
Double cover of $T^\star(\mathbb{CP}^1)$ is also introduced and 
its topological stability 
is argued. 
In sect.\ref{sc:SUSYbreaking}, 
SUSY breaking exhibited as 
mass splitting between light bosons and 
fermions is discussed. 
Useful formulas of gamma matrices and spinors in 
four and five dimensions are summarized in 
Appendix \ref{ap:gamma}. 
Massive modes for one of the field 
in spherical coordinates is worked out in 
Appendix \ref{ap:omega-mode}.

\section{The non-BPS domain walls in five dimensions}
\label{sc:nonBPSsolution}
\subsection{The BPS and non-BPS solutions in the sine-Gordon model}
In this subsection we briefly review BPS and non-BPS domain 
walls in an ${\cal N}=1$ SUSY complex sine-Gordon model 
in four dimensions
\cite{Maru:2000sx,Maru:2001gf}.
It contains a chiral superfield 
\begin{eqnarray}
A(y^m,\theta) = a(y) + \sqrt 2 \theta \psi(y) + \theta^2 F(y),\quad
y^m \equiv x^m + i\theta\sigma^m\bar\theta, 
\end{eqnarray}
with 
the sine-Gordon superpotential $P$ and 
with the minimal kinetic term 
\begin{eqnarray}
P(A) = \frac{\Lambda^3}{g^2}\sin\frac{g}{\Lambda}A,\quad
K(A,\bar A) = \bar AA,
\end{eqnarray}
where $K$ is the 
K\"ahler potential, 
$\Lambda$ is a coupling constant of unit mass 
dimension, and $g$ is a dimensionless coupling constant. 
The spacetime index $m$ runs from 0 to 3.
The bosonic part of the Lagrangian reads 
\begin{eqnarray}
{\cal L}_{\rm boson} &=&
- |\partial_ma|^2 - \frac{\Lambda^4}{g^2} 
\left|\cos\frac{g}{\Lambda}a\right|^2,
\end{eqnarray}
where the auxiliary fields $F$ is eliminated. 
Defining dimensionless real scalar fields $(\Theta,\Phi)$ 
\begin{eqnarray}
\Theta - \frac{\pi}{2} + i\Phi \equiv 
\frac{g}{\Lambda}\left({\rm Re}[a] + i{\rm Im}[a]\right), 
\end{eqnarray}
we can rewrite the above Lagrangian as 
\begin{eqnarray}
{\cal L}_{\rm boson} = \frac{2\Lambda^2}{g^2}\left[
- \frac{1}{2}\partial_m\Theta \partial^m\Theta
- \frac{1}{2}\partial_m\Phi \partial^m\Phi
- \frac{\Lambda^2}{2}\sin^2\Theta
- \frac{\Lambda^2}{2}\sinh^2\Phi\right].
\label{eq:sine-Gordon-Lag}
\end{eqnarray}
This model has infinitely many isolated SUSY vacua at 
$\Theta = n\pi,\ \Phi=0$ $(n\in Z)$. 
The existence of two or more isolated vacua can admit 
domain wall solutions interpolating between these vacua. 
The variable $\Theta$ may be regarded as taking 
any real values. 
However, the Lagrangian (\ref{eq:sine-Gordon-Lag}) 
with the sine-Gordon superpotential has the periodicity 
in $\Theta \simeq \Theta + 2\pi$. 
Therefore the variable $\Theta$ is naturally a periodic 
variable taking values in $\Theta \in [0,2\pi)$. 
On the other hand, $\Phi$ has no periodicity. 
Then the target space of the Lagrangian 
(\ref{eq:sine-Gordon-Lag}) is 
$S^1\times \mathbb{R}$. 
In that sense, there are only two isolated vacua 
at $\Theta = 0,\pi,\ \Phi=0$ in the fundamental 
domain ($0\le \Theta < 2\pi,  -\infty \le \Phi <  \infty$). 

Let us assume that the wall 
has a nontrivial profile in the $y$ coordinate which 
is identified as the extra dimension.  
The energy density (tension) of the domain wall is 
bounded by the Bogomolny bound\cite{Bogomolny:1975de} 
\begin{eqnarray}
E 
&=& \frac{2\Lambda^2}{g^2}\int_{-\infty}^{\infty} dy\ 
\bigg[\frac{1}{2}\left(\Theta' \mp \Lambda 
\sin\Theta \cosh\Phi\right)^2
+ \frac{1}{2}\left(\Phi' \pm 
\Lambda\cos\Theta\sinh\Phi\right)^2\nonumber\\
&&\qquad \pm \Lambda\left(\Theta'\sin\Theta\cosh\Phi  
- \Phi'\cos\Theta\sinh\Phi\right)
\bigg]\nonumber\\
&\ge& \frac{2\Lambda^2}{g^2}
\left[\mp \Lambda \cos \Theta 
\cosh\Phi \right]^\infty_{-\infty},
\label{eq:tension}
\end{eqnarray}
where prime denotes the derivative with respect to $y$. 
The Bogomolny bound is saturated when the 
following BPS equations are satisfied : 
\begin{eqnarray}
\Theta' = \pm \Lambda \sin \Theta \cosh \Phi,\quad
\Phi' = \mp \Lambda \cos \Theta \sinh\Phi.
\label{BPS_sine-Gordon}
\end{eqnarray}
Imposing the boundary condition such as 
$(\Theta,\Phi) = (0,0)$ or $(\pi,0)$ at 
minus (plus) infinity of $y$, we find 
the above BPS equation becomes simpler 
$\Theta' = \pm \Lambda \sin\Theta,\ \Phi = 0$
leading to the BPS single wall solutions 
with $y_0$ as the moduli parameter associated 
with the center of the mass position of the 
domain wall\cite{Maru:2001gf}:
\begin{eqnarray}
\Theta(y;y_0) = 
\pm \sin^{-1}\left(\tanh\left(\Lambda(y-y_0)\right)\right) 
+ \frac{\pi}{2}, 
\label{eq:BPS-wall}
\end{eqnarray}
whose tension is given by $E=\frac{4\Lambda^3}{g^2}$ 
from the boundary conditions and 
Eq.(\ref{eq:tension}). 
The half of SUSY charges are preserved by 
the BPS wall solution with plus sign in 
Eq.(\ref{eq:BPS-wall}), 
whereas the other half is preserved 
by the BPS solution with the minus sign. 
To distinguish the preserved SUSY charges, 
we shall call the solution with the minus (plus) 
sign as anti-BPS (BPS) solution. 

\begin{figure}[ht]
\begin{center}
\includegraphics{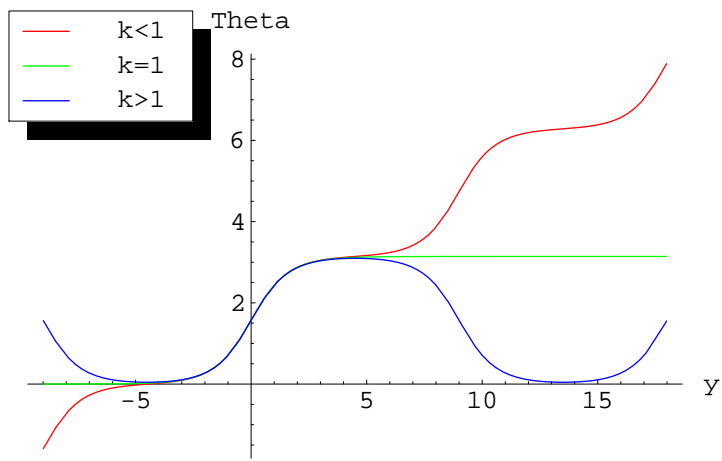}
\end{center}
\caption{The Jacobi's amplitude function 
with $k>1,\ k= 1,\ k<1$.}
\label{am_k}
\end{figure}

Let us next consider non-BPS domain wall solutions 
of the equations of motion. 
Since the potential monotonically increases 
as ${\rm e}^{|\Phi|}$ for 
$|\Phi|\rightarrow \infty$, 
we should 
look for 
solutions with $\Phi=0$. 
Then 
the Lagrangian and the equations of motion reduce to
\begin{eqnarray}
{\cal L}_{\rm boson} &=& \frac{\Lambda^2}{g^2}
\left[ - \frac{1}{2}(\Theta')^2 
- \frac{\Lambda^2}{2}\sin^2\Theta \right],
\\
\Theta'' &=& \Lambda^2\sin\Theta\cos\Theta.
\label{eq:EOM-sineGordon}
\end{eqnarray}
An exact solution of this 
equation has been found\cite{Maru:2001gf} 
with two parameters $y_0, k$
\begin{eqnarray}
\Theta(y;y_0,k) = 
{\rm am}\left(\frac{\Lambda}{k} (y-y_0),k \right) 
+ \frac{\pi}{2},
\qquad 0< k, 
\label{nbps_sine}
\end{eqnarray}
where  the amplitude function ${\rm am}(u,k)$
is defined in terms of the Jacobi's elliptic 
function ${\rm sn}(u,k)$ as 
${\rm am}(u,k) = \sin^{-1}{\rm sn}(u,k)$. 
The elliptic functions 
${\rm sn}(u, k), {\rm cn}(u, k)$ are 
periodic in $u$ with the period of 
$4K(k)$, where $K$ is the complete elliptic integral 
of the first kind. 
Therefore we can compactify the base space with 
the radius $L$ by requiring
\footnote{
Alternative choices are 
$2n\pi L = 4 k K(k)/\Lambda, \; n=1, 2, \cdots$, 
corresponding to $n$ pairs of BPS wall and anti-BPS wall 
placed with equal interval in the fundamental region 
$2\pi L$. 
} 
$2\pi L = 4 k K(k)/\Lambda$. 
When $k=1$, the radius diverges 
($L\rightarrow \infty$) and the solution 
reduces to the BPS 
wall solution in Eq.(\ref{eq:BPS-wall}) 
with $y_0$ as the wall position. 
For $k \not =1$, the solution breaks SUSY 
completely, giving a non-BPS two wall solution 
with BPS and anti-BPS walls situated at $y_0$ 
and $y_0+\pi L$. 
For $k<1$, the solution is quasi-periodic and 
represents a non-BPS two walls with unit winding 
number in the target space. 
For $k>1$, the solution is periodic and represents 
wall and anti-wall with no winding. 
These three cases of non-BPS solutions 
are illustrated in Fig.\ref{am_k}.

Since the non-BPS solution (\ref{nbps_sine}) 
breaks SUSY completely, 
its stability is not ensured by 
the central charge of the SUSY algebra. 
In fact it has been found that the non-BPS solutions 
for $k>1$ is unstable with tachyon, and 
those for $k<1$ is stable because of the nontrivial 
winding number\cite{Maru:2001gf}. 
To see this point, one examines 
small fluctuations $\theta(x,y),\varphi(x,y)$ 
around the non-BPS background 
configurations in Eq.(\ref{nbps_sine}). 
The linearized 
equations of motion for the fluctuation fields 
are given by 
\begin{eqnarray}
\partial_\mu\partial^\mu \theta &=& \left[- \frac{d^2}{dy^2} 
+ \Lambda^2\left(2{\rm sn}^2
\left(\frac{\Lambda}{k}(y-y_0),k\right) - 1\right)
\right]\theta,\label{seq_theta}\\
\partial_\mu\partial^\mu \varphi 
&=& \left[- \frac{d^2}{dy^2} + \Lambda^2\right]\varphi,
\label{seq_phi}
\end{eqnarray}
where the world volume coordinates are 
denoted by $x^\mu$ with $\mu$ 
running from 0 to 2. 
Mode functions $\psi_\theta^{(l)}(y)$ 
($\psi_\varphi^{(l)}(y)$) 
for fluctuation field 
$\theta$ ($\varphi$) are defined by 
\begin{eqnarray}
&&\left[ - \frac{d^2}{dy^2} + \Lambda^2\left(
2{\rm sn}^2\left(\frac{\Lambda}{k}(y-y_0),k\right)-1\right)
\right]\psi_\theta^{(l)}
= (m_{\theta}^{(l)})^2\psi_\theta^{(l)},
\label{eq:eigen-theta}
\\
&&\left[ - \frac{d^2}{dy^2} + \Lambda^2\right]\psi_\varphi^{(l)}
= (m_{\varphi}^{(l)})^2 \psi_\varphi^{(l)}.
\label{eq:eigen-varphi}
\end{eqnarray}
One can expand the fluctuation fields in terms of 
these mode functions yielding effective fields 
$f_\theta^{(m)}(x)$ ($f_\varphi^{(m)}(x)$) on 
the world volume 
with mass squared $(m_{\theta}^{(l)})^2$ ($(m_{\varphi}^{(l)})^2$) 
\begin{eqnarray}
\theta(x,y) = \sum_m \psi_\theta^{(m)}(y) f_\theta^{(m)}(x),\quad
\varphi(x,y) = \sum_m \psi_\varphi^{(m)}(y) f_\varphi^{(m)}(x).
\end{eqnarray}

Since the Schr\"odinger-like eigenvalue equation 
(\ref{eq:eigen-varphi}) for 
$\varphi$ obviously has no negative eigenvalues, 
$\varphi$ does not have a tachyon. 
One can obtain three lowest eigenmodes 
exactly\cite{Maru:2001gf} for the eigenvalue equation 
(\ref{eq:eigen-theta}) for $\theta$ : 
\begin{alignat}{2}
&m_{\theta,0}^2 = 0,\quad 
&\psi_\theta^{(0)} = {\rm dn}\left(\frac{\Lambda}{k}(y-y_0),k\right),\\
&m_{\theta,1}^2 = \frac{1-k^2}{k^2}\Lambda^2,\quad 
&\psi_\theta^{(1)} = {\rm cn}\left(\frac{\Lambda}{k}(y-y_0),k\right),\\
&m_{\theta,2}^2 = \frac{1}{k^2}\Lambda^2,\quad 
&\psi_\theta^{(2)} = {\rm sn}\left(\frac{\Lambda}{k}(y-y_0),k\right). 
\end{alignat}

\begin{figure}[ht]
\begin{center}
\includegraphics{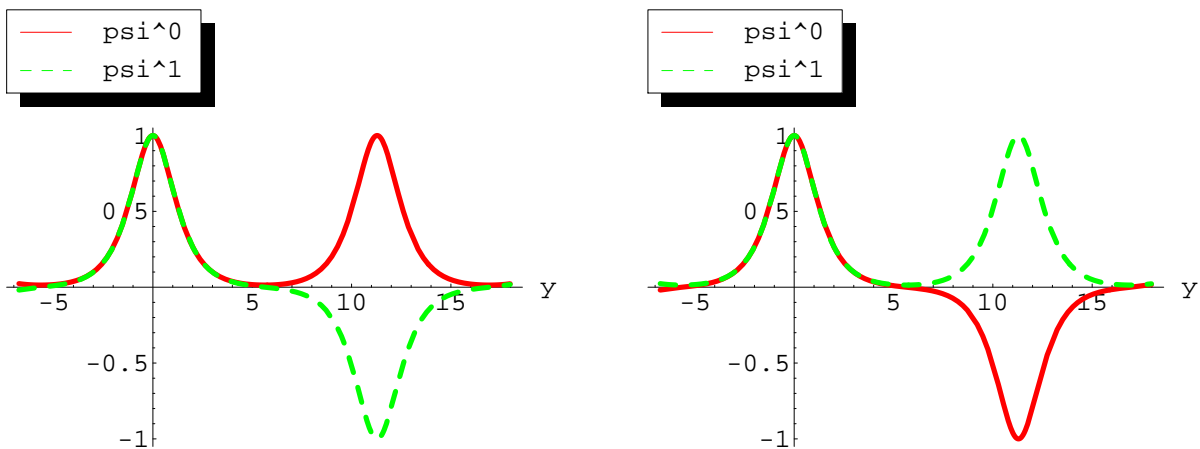}
\caption{{\small The wave functions of $\theta$ are depicted.
The red line denotes translational zero mode $\psi_\theta^{(0)}$ and
the green line denotes breather mode $\psi_\theta^{(1)}$. The left hand side corresponds to
the case of $k<1$ and the right hand side to $k>1$.}}
\label{wavefunc}
\end{center}
\end{figure}

The massless mode $\psi_\theta^{(0)}$ is 
the Nambu-Goldstone 
mode for the broken translational invariance. 
In fact, its profile in the left part of 
Fig.\ref{wavefunc} 
is positive definite corresponding to 
the derivative of the monotonically increasing 
background configuration for $k<1$. 
As shown in the left part of 
Fig.\ref{wavefunc}, 
the first excited mode 
$\psi_\theta^{(1)}$ has a profile 
of the difference between 
the translational zero-modes of individual 
BPS wall and anti-BPS wall. 
Therefore it corresponds to 
the fluctuation of relative distance between two walls, 
so-called breather mode. 
For the case of $k<1$, the mass squared of 
the breather mode $\psi_\theta^{(1)}$ 
is positive, showing 
the stability of the background non-BPS configuration 
with nonzero winding number. 
For the case of $k>1$, 
on the contrary, 
$\psi_\theta^{(1)}$ 
becomes tachyon destabilizing the background without 
winding number. 
Since the radius diverges in the limit 
$k\rightarrow 1$, the 
anti-BPS wall at $y=y_0+\pi L$ goes to infinity, and the solution 
reduces to the BPS single wall solution. 
In this limit, $\psi_\theta^{(1)}$ becomes massless, 
and the sum of $\psi_\theta^{(1)}$ and $\psi_\theta^{(0)}$ 
is localized on the BPS wall, whereas the difference 
is localized on the anti-BPS wall, which disappears to 
infinity. 

\subsection{BPS and non-BPS Domain walls in five dimensions}
\label{sc:walls-infinite-coupl}
\subsubsection{Models admitting domain walls}
Although a stable non-BPS solutions with 
a winding number has been obtained for a 
model in four dimensions, we need a model in five 
spacetime dimensions to build a realistic brane-world
by thick walls. 
The models should have discrete SUSY vacua for 
domain walls. 
This ca be achieved either 
by a SUSY gauge theories interacting with hypermultiplets, 
or by nonlinear sigma models of hypermultiplets. 
As a gauge theory, one can take a SUSY $U(N_c)$ gauged 
theory with $N_f>N_c$ flavors of 
hypermultiplets in the fundamental representation. 
If the hypermultiplet  masses are nondegenerate 
and the $U(1)$ factor group of $U(N_c)$ has the 
Fayet-Iliopoulos (FI) terms, the model exhibits 
discrete SUSY vacua~\cite{AraiNittaSakai}. 

For simplicity we will consider the SUSY $U(1)$ gauge 
theory with $N_f (\ge 2)$ hypermultiplets.
The vector multiplet consists of 
a five-dimensional gauge field $A_M$, 
a symplectic Majorana fermion $\Lambda^i$ which 
satisfies the symplectic Majorana condition 
$\Lambda^i = \varepsilon^{ij} C \bar\Lambda_j^T$ 
in five dimensions
\footnote{The conventions of gamma matrices 
and the spinors in five dimensions are 
given in Appendix \ref{ap:gamma}.} 
and a real adjoint (neutral) scalar field $\Sigma$.
The hypermultiplets 
consist of charged scalar fields $H_A^i$ 
and their fermionic superpartners $\Psi_A$. 
The $SU(2)_R$ doublet index $i$ runs 
$i=1, 2$ 
and the 
flavor index $A$ runs from 1 to $N_f$. 
Notice that the real degrees of freedom is 
eight both for 
the symplectic Majorana fermion $\Lambda^i$ 
and for 
the Dirac spinor $\Psi_A$ in five dimensions. 
We denote the gauge coupling by $e$. 
The Lagrangian is given by 
\begin{eqnarray}
{\cal L} &=& 
- \frac{1}{4e^2} F^{MN}F_{MN} - \frac{1}{2e^2} 
\partial_M\Sigma\partial^M\Sigma 
- \sum_{A=1}^{N_f}{\cal D}^MH_{Ai}^* {\cal D}_M H_A^i\nonumber\\
&& - \frac{i}{2e^2}\bar\Lambda_i\Gamma^M\partial_M\Lambda^i
- \frac{i}{e^2} \sum_{A=1}^{N_f} \bar\Psi_A\Gamma^M
{\cal D}_M\Psi_A\nonumber\\
&&+ \sum_{A=1}^{N_f}
\left[i \sqrt 2 \varepsilon_{ij}\bar\Psi_A\Lambda^iH_A^j
 - i \sqrt 2 \varepsilon^{ij}H_{Ai}^*\bar\Lambda_j \Psi_A
- (\Sigma - \mu_A)\bar\Psi_A\Psi_A\right] - V,\\
V &=& \frac{e^2}{2}\sum_{a=1}^3
\left( -2\xi\delta_3^a + 
\sum_{A=1}^{N_f} H^*_{Ai}(\sigma^a)^i{}_jH_A^j\right)^2
+ \sum_{A=1}^{N_f}\left(\Sigma + \mu_A\right)^2 H_{Ai}^*H_A^i, 
\label{scalar_pot}
\end{eqnarray}
where $\mu_A$ is the mass of the $A$-th hypermultiplet 
and the covariant derivatives are 
\begin{eqnarray}
{\cal D}_M H_A^i 
=
\left(\partial_M + iA_M\right)H_A^i,
\qquad 
{\cal D}_M \Psi_A 
=
 \left(\partial_M + iA_M\right)\Psi_A.
\end{eqnarray}
The $SU(2)_R$ triplet FI parameters 
are chosen to lie in the third direction and is 
denoted as $\xi$. 

Let us first examine SUSY vacua. 
We denote the $SU(2)_R$ components of 
the hypermultiplets as 
\begin{eqnarray}
H_A^i = \left(
\begin{array}{c}
H_A\\
H_A^{c*}
\end{array}
\right),\quad
H_{Ai}^\dagger
= \left(
\begin{array}{cc}
H_A^* & H_A^c
\end{array}
\right).
\end{eqnarray}
Choosing nondegenerate mass parameters $\mu_A\not=\mu_B$, 
we obtain $N_f$ SUSY vacua in the Higgs phase. 
The $A$-the vacuum is given by 
\begin{eqnarray}
\Sigma = - \mu_A,\quad
H_B^c = 0,\quad
|H_B|^2 = 2\xi\delta^A_B\quad
(B=1,2,\cdots,N_f).
\end{eqnarray}
Therefore we expect the existence of (multi) BPS domain 
walls which interpolates a pair 
of these discrete Higgs vacua. 
The minimal model admitting such a 
BPS domain 
wall is the case of $N_f=2$, which will be considered 
from now on.

Even with this simple model, it is generally 
difficult to obtain exact wall solutions for the case 
of finite gauge coupling~\cite{KakimotoSakai}, 
\cite{Isozumi:2003rp}. 
Although we will consider also finite gauge coupling $e$ 
later, it is sufficient to examine 
the case of infinite 
gauge coupling to study domain walls. 
We will see that we 
can obtain exact solutions 
in the infinite gauge coupling limit 
not only for 
BPS single wall configurations but 
also for non-BPS multi-wall configurations.
As we let gauge coupling to infinity 
$e\rightarrow \infty$, 
the kinetic term of vector multiplet in the Lagrangian 
vanishes. 
At the same time, the scalar potential becomes infinitely 
steep and the hypermultiplets are constrained to be at 
the minimum 
\begin{eqnarray}
\sum_{A=1}^{N_f} H_A^cH_A 
= \sum_{A=1}^{N_f} H_A^*H_A^{c*} = 0,\quad
\sum_{A=1}^{N_f}\left( |H_A|^2 - |H_A^c|^2 \right) 
= 2\xi.\label{T^*CP^1}
\end{eqnarray}

The gauge field $A_M$ and the adjoint scalar field 
$\Sigma$ in the vector multiplet become Lagrange 
multiplier fields which can be eliminated 
to give the reduced Lagrangian ${\cal L}_\infty$ 
at infinite coupling $e\rightarrow \infty$ 
\begin{eqnarray}
{\cal L}_\infty 
&=& - \sum_{A=1}^{N_f}
\left(|\partial_MH_A|^2 + |\partial^MH_A^c|^2\right) 
+ \frac{\left[\sum_{A=1}^{N_f}
\left(H_A^*\overleftrightarrow\partial_{\!\!\!M}H_A 
+ H_A^c\overleftrightarrow
\partial_{\!\!\!M}H_A^{c*}\right)\right]^2}{
4\sum_{A=1}^{N_f}\left(|H_A|^2 + |H_A^c|^2\right)}\nonumber\\
&& - \sum_{A=1}^{N_f}
\left(\mu_A^2 |H_A|^2 + \mu_A^2 |H_A^c|^2\right) 
+ \frac{\left[\sum_{A=1}^{N_f}
\left(\mu_A |H_A|^2 + \mu_A|H_A^c|^2\right)\right]^2}{
4\sum_{A=1}^{N_f}\left(|H_A|^2 + |H_A^c|^2\right)},
\label{eq:Lag-inf}
\end{eqnarray}
where we denote 
$X\overleftrightarrow\partial_{\!\!\!M}Y\equiv 
X\partial_MY - Y\partial_MX
$. 
Taking the infinite gauge coupling limit 
$e\rightarrow \infty$ of 
the SUSY gauge theory with massive hypermultiplets 
gives a nonlinear sigma model with a potential 
term as seen above. 
This is called the massive hyper K\"ahler 
quotient method\cite{RT}, 
\cite{LR}, \cite{Arai:2002xa}, \cite{Arai:2003es}.

The simplest model with $N_f=2$ hypermultiplets is a 
nonlinear sigma model with $T^\star(\mathbb{CP}^1)$ as 
target space and with an appropriate potential, 
which is called the massive $T^\star(\mathbb{CP}^1)$ model. 
To solve the constraint for hypermultiplet scalars in 
Eq.(\ref{T^*CP^1}), we introduce spherical coordinates 
$(R, \Omega, \Theta, \Phi)$ as four independent variables
\cite{CF,Arai:2003es,Arai:2002xa,Alvarez-Gaume:1983ab,
Gibbons:1987pk} 
\begin{eqnarray}
  H_1 
&
=
&
g(R)\cos\left(\frac{\Theta}{2}\right)
\exp\left(\frac{i}{2}(\Omega+\Phi)\right), 
\\
H_2 
&
=
&
g(R)\sin\left(\frac{\Theta}{2}\right)
\exp\left(\frac{i}{2}(\Omega-\Phi)\right), 
\\
H_1^{c*} 
&
=
&
f(R)\sin\left(\frac{\Theta}{2}\right)
\exp\left(-\frac{i}{2}(\Omega-\Phi)\right), 
\\
H_2^{c*} 
&
=
&
-f(R)\cos\left(\frac{\Theta}{2}\right)
\exp\left(-\frac{i}{2}(\Omega+\Phi)\right) .
\end{eqnarray}
where $f(R)$ and $g(R)$ are given by 
\begin{eqnarray}
f(R)^2 = - \xi + \sqrt{R^2 + \xi^2},\quad
g(R)^2 = \xi + \sqrt{R^2 + \xi^2}.
\end{eqnarray}
The range of these variables are usually taken as 
\begin{eqnarray}
0 \le R < \infty,\quad
0 \le \Theta \le \pi,\quad
0 \le \Phi \le 2\pi,\quad
0 \le \Omega \le 2\pi.
\end{eqnarray} 
This is one of the standard parametrizations 
of $T^\star(\mathbb{CP}^1)$ manifold, which is also called 
Eguchi-Hanson manifold. 
In terms of these independent 
variables $(R,\Theta,\Omega,\Phi)$, 
the Lagrangian (\ref{eq:Lag-inf}) reads 
\begin{eqnarray}
{\cal L}_\infty &=&
\frac{1}{2\sqrt{R^2 + \xi^2}}\bigg[
- \partial^MR \partial_M R - (R^2 + \xi^2)\partial_M
\Theta\partial^M\Theta
- \left(R^2 + \xi^2\sin^2\Theta\right)\partial_M
\Phi\partial^M\Phi\nonumber\\
&&- R^2\partial_M\Omega\partial^M\Omega 
- 2R^2\cos\Theta\partial_M\Phi\partial^M\Omega
- \mu^2\left(R^2 + \xi^2\sin^2\Theta\right)\bigg].
\label{EH_lagrangian}
\end{eqnarray}
Since a common mass of hypermultiplets can be absorbed 
by a shift of vector multiplet scalar $\Sigma$, we 
set $\mu_1 = -\mu_2 = \frac{\mu}{2}$ here. 

\begin{figure}[ht]
\begin{center}
\includegraphics{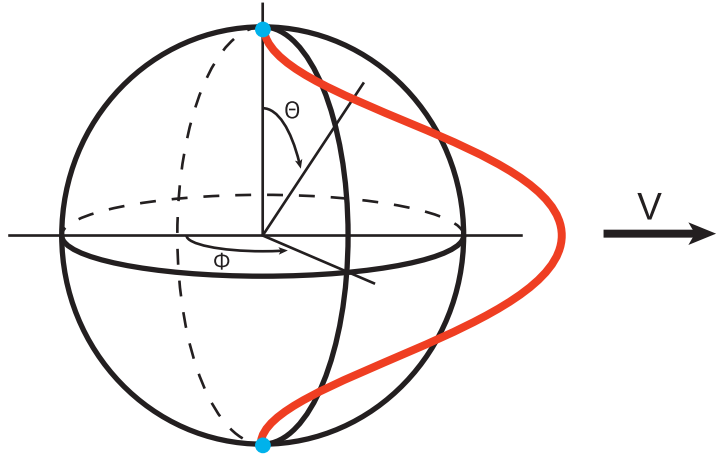}
\end{center}
\caption{The target space of truncated Eguchi-Hanson 
manifold and the scalar potential
on the manifold. The isolated SUSY vacua are at the 
north pole and south pole.}
\label{s2}
\end{figure}

Let us notice that  $(R, \Omega)$ parametrize the fiber 
of $T^\star(\mathbb{CP}^1)$ and 
the submanifold 
defined by $(R=0,\Omega=0)$ is the base 
space $\mathbb{CP}^1$. 
If we truncate the manifold to the base manifold, 
it is a K\"ahler manifold $\mathbb{CP}^1$, which is 
just a sphere $S^2$ with the radius $\xi$. 
Two coordinates $\Theta$ and $\Phi$ correspond 
to the latitude and the longitude of 
the sphere, as illustrated in Fig.\ref{s2}.
The scalar potential on the sphere is given by
\begin{eqnarray}
V_{CP^1} = \xi \frac{\mu^2}{2}\sin^2\Theta, 
\end{eqnarray}
which has two isolated SUSY vacua at the north 
and south pole of the sphere.
\subsubsection{The BPS and non-BPS domain walls in five dimensions}
In this subsection we obtain BPS and non-BPS 
domain wall configurations 
in the nonlinear sigma model $T^\star(\mathbb{CP}^1)$.
Similarly to the sine-Gordon model, the Bogomolny 
bound can be obtained:
\begin{eqnarray}
E 
&=& \int dy\ 
\frac{1}{2\sqrt{R^2+\xi^2}}\bigg[
\left(R^2 + \xi^2\right)
\left(\Theta' \pm \mu\sin\Theta\right)^2
+ \left(R^2+\xi^2\right)\Phi'{^2}
\sin^2\Theta
\nonumber\\
&&\qquad
+ \left(R'\mp\mu R \cos\Theta\right)^2
+ R^2\left(\Omega' + \Phi'\cos\Theta\right)^2\nonumber\\
&&\qquad
\pm 2\mu R R'\cos\Theta \mp 2 \mu 
\left(R^2 + \xi^2\right)\Theta'\sin\Theta
\bigg]\nonumber\\
&\ge&
\pm \mu \left[\sqrt{R^2 + \xi^2}
\cos\Theta\right]^{\infty}_{-\infty}.
\end{eqnarray}
This energy bound is saturated when the following 
BPS equations are satisfied:
\begin{eqnarray}
\Theta' = \pm \mu \sin\Theta,&&\quad
\Phi' \sin\Theta = 0,
\label{BPSeq:Theta}
\\
R' = \mp \mu R \cos \Theta,&&\quad
R\left(\Omega' + \Phi'\cos\Theta\right) = 0,
\label{BPSeq:R}
\end{eqnarray}
where we assume the background configuration to 
depend only on the extra dimension coordinate $y$. 
Since we are now interested in BPS wall 
solutions interpolating between 
two SUSY vacua with $R=0, \Omega=0$, 
we assume $R=0,\ \Omega=0$ for the wall 
configurations. 
Then the BPS equations 
reduce to
\begin{eqnarray}
\Theta' = \pm \mu \sin\Theta,\quad
\Phi' \cdot \sin\Theta = 0.
\end{eqnarray}
From the second equation, $\Phi$ must be a constant.
Thus the BPS solution which interpolates the two 
isolated vacua at 
the north pole and south pole of $S^2$, is on a 
great circle $\Phi = {\rm const.}$.
Notice that this BPS equation for $\Theta$ is 
identical to the BPS equation 
(\ref{BPS_sine-Gordon}) for $\Theta$ in the 
sine-Gordon model.
Therefore we obtain the BPS wall solution :
\begin{eqnarray}
\Theta(y;y_0) = \pm \sin^{-1}\left(\tanh\mu(y-y_0) \right)
+ \frac{\pi}{2},\quad
\Phi = {\rm const.},
\label{eq:BPSsol-CP1}
\end{eqnarray}
where $y_0$ is the position of the wall. 

Now we will look for the non-BPS solutions,  
consisting of a BPS wall and an anti-BPS wall. 
Inspired by the BPS wall solutions, 
let us consider solutions of equations of motion 
with vanishing values and derivatives for $R, \Omega$ 
as their initial conditions at some $y$. 
Then the equations of motion dictates $R=\Omega=0$ 
for all values of $y$. 
Therefore we can truncate the 
$T^\star(\mathbb{CP}^1)$ model, and 
can consider $\mathbb{CP}^1$ model effectively 
\begin{eqnarray}
{\cal L}^{CP^1}_{\rm boson} = 
\frac{\xi}{2}\left( - \partial_M\Theta\partial^M\Theta 
- \sin^2\Theta\partial_M\Phi\partial^M\Phi
- \mu^2 \sin^2\Theta\right).
\end{eqnarray}
\begin{figure}[t]
\begin{center}
\includegraphics[width=15cm]{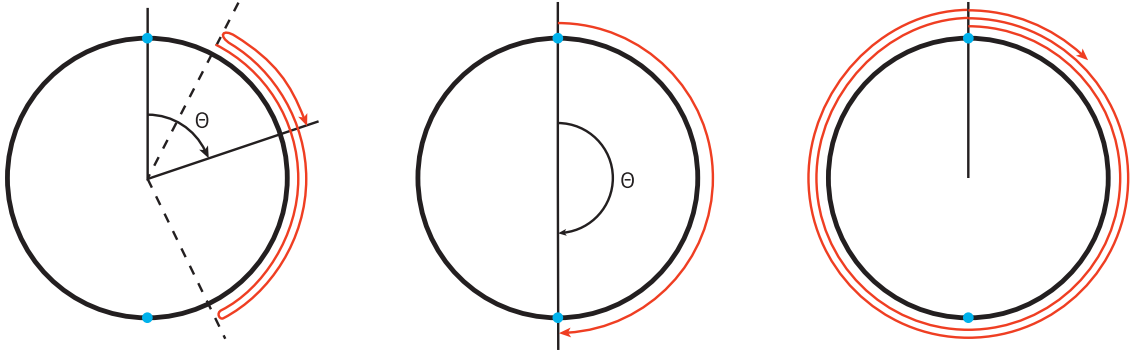}
\caption{The non-BPS wall solutions are shown. 
The left figure shows 
the periodic solution with $k>1$, the middle 
shows BPS or anti-BPS solution 
with $k=1$ 
and the right shows quasi-periodic solution with $k<1$.
}
\label{nBPS_cp1}
\end{center}
\end{figure}
Assuming $\Theta$ and $\Phi$ to depend on $y$ only, 
their equations of motion becomes :
\begin{eqnarray}
&&\Theta'' - \left({\Phi'}^2 + \mu^2\right)
\sin\Theta\cos\Theta = 0,\\
&&\left[\sin^2\Theta \Phi'\right]' = 0.
\end{eqnarray}
Similarly to $R, \Omega$, we can consider initial 
conditions $\Phi'=0$. 
Then $\Phi$ becomes constant, and the 
equation of motion for $\Theta$ reduces to
\begin{eqnarray}
\Theta'' = \mu^2 \sin\Theta\cos\Theta.
\label{eq:BPS-eq-theta}
\end{eqnarray}
This is identical to the equation of motion for 
the real part of the sine-Gordon model in 
Eq.(\ref{eq:EOM-sineGordon}). 
Therefore we obtain a non-BPS solution 
\begin{eqnarray}
\Theta(y;y_0,k) 
= {\rm am}\left(\frac{\mu}{k}(y-y_0),k\right) 
+ \frac{\pi}{2},\quad
\Phi = {\rm const.}.
\label{nbps_cp1}
\end{eqnarray}
We can compactify the base space by 
$2\pi L = 4k K(k)/\mu$, similarly to 
the sine-Gordon model\footnote{
Alternative choices are 
$2n\pi L = 4 k K(k)/\Lambda, \; n=1, 2, \cdots$, 
corresponding to $n$ pairs of BPS wall and anti-BPS wall 
placed with equal interval in the fundamental region 
$2\pi L$. 
} 
. 
Besides $y_0$ representing the position of the wall, 
this solution has one more parameter $k$. 
In the case of $k>1$ the solution curve never reaches 
either vacuum at north and south poles, and oscillates 
in an interval between them. 
In the case of $k=1$ the solution corresponds to
the BPS or anti-BPS solution in Eq.(\ref{eq:BPSsol-CP1}) 
which interpolates north pole and south pole once.
In the case of $k<1$ the solution passes through 
both vacua and becomes quasi-periodic. 
Similarly to the sine-Gordon case, this solution 
represents the BPS wall and the anti-BPS wall 
placed at $y_0$ and at $y_0+\pi L$, respectively. 
These are depicted in Fig.\ref{nBPS_cp1}.
The solution and the model have similarities and 
differences 
with those of the sine-Gordon model in four dimensions, as 
listed in Table\ref{table:sG-TCP1}.  
\begin{table}[t]
\centering
\caption{Correspondence between the sine-Gordon model 
in four dimensions 
and $T^\star(\mathbb{CP}^1)$ model in five dimensions.}
\label{table:sG-TCP1}
\begin{tabular}{c|c}
sine-Gordon model & $T^\star(\mathbb{CP}^1)$ model\\
\hline
${\rm Re}(a) 
= \frac{\Lambda}{g}{\rm am}\left(\frac{\Lambda}{k}(y-y_0),k\right)$ & 
$\Theta 
= {\rm am}\left(\frac{\mu}{k}(y-y_0),k\right) 
+ \frac{\pi}{2}$\\
${\rm Im}(a) = 0$ & $\Phi = {\rm const.}$\\
nothing & $R=0$\\
nothing & $\Omega=0$
\end{tabular}
\end{table}
In terms of the original hypermultiplet variables 
$H_A,H^c_A$, the above non-BPS solution is 
expressed as 
\begin{eqnarray}
H_1 &=& \sqrt{2\xi} {\rm e}^{i\frac{\Phi}{2}} \cos 
\left[\frac{1}{2}{\rm am}
\left(\frac{\mu}{k}(y-y_0),k\right) + \frac{\pi}{4}\right],\\
H_2 &=& \sqrt{2\xi} {\rm e}^{-i\frac{\Phi}{2}} \sin
\left[\frac{1}{2}{\rm am}
\left(\frac{\mu}{k}(y-y_0),k\right) + \frac{\pi}{4}\right],\\
H_1^c &=& H_2^c = 0.
\end{eqnarray}

\subsection{Domain walls in the finite gauge coupling}

Here we review BPS solutions at finite gauge 
coupling~\cite{Isozumi:2003rp}, 
and examine the non-BPS case also, 
to obtain an idea of how the domain wall solutions 
are modified at finite gauge 
coupling. 
The bosonic Lagrangian with the finite gauge coupling 
is given by
\begin{eqnarray}
{\cal L}_{\rm boson} \!\!
&=&\!\! - \frac{1}{4e^2} F^{MN} F_{MN} 
- \frac{1}{2e^2} \partial^M \Sigma \partial_M \Sigma
- \sum_A\left(|{\cal D}^MH_A|^2 
- |{\cal D}^MH_A^c|^2\right) - V,\\
V \!\!&=&\!\! 2e^2\left|\sum_A H_AH_A^c\right|^2
+ \frac{e^2}{2} \left( \sum_A |H_A|^2 
- |H_A^c|^2 - 2\xi \right)^2\nonumber\\
&&\qquad+ \sum_A(\Sigma + \mu_A)^2 
\left( |H_A|^2 + |H_A^c|^2 \right).
\end{eqnarray}
Similarly to the infinite coupling case, 
BPS wall solution should be obtained with 
$H^c = 0$ and $A_M = 0$, since it 
interpolates between two SUSY vacua 
with $H^c=0$ respecting the four-dimensional 
Poincar\'e invariance. 
Assuming that the scalars $H_1$ and $H_2$ depend 
on $y$ only, we find that the above Lagrangian reduces to 
\begin{eqnarray}
{\cal L}_{\rm boson} &=& - \frac{1}{2e^2}(\Sigma')^2
- |H_1'|^2 - |H_2'|^2 - V,\\
V &=& \frac{e^2}{2}\left(|H_1|^2 + |H_2|^2 - 2\xi\right)^2
+ \left(\Sigma + \frac{\mu}{2}\right)^2|H_1|^2 
+ \left(\Sigma - \frac{\mu}{2}\right)^2|H_2|^2.
\end{eqnarray}
The tension of the domain wall is given by 
\begin{eqnarray}
E &=& \int dy\ 
\left[
\frac{1}{2e^2}(\Sigma')^2
+ |H_1'|^2 + |H_2'|^2 + V
\right]\nonumber\\
&=& \int dy\ 
\left[
\frac{1}{2e^2}\left\{ \Sigma' 
+ e^2\left(|H_1|^2 + |H_2|^2 - 2\xi\right) \right\}^2
+ \left| H_1' 
+ \left(\Sigma + \frac{\mu}{2}\right) H_1 \right|^2
\right] \nonumber\\
&+& \int dy\ 
\left[ 
\left| H_2' + \left(\Sigma 
- \frac{\mu}{2}\right) H_2 \right|^2
+ \left\{
\left(\Sigma + \frac{\mu}{2}\right)|H_1|^2 
+ \left(\Sigma - \frac{\mu}{2}\right)|H_2|^2 
- 2\xi\Sigma \right\}'
\right]\nonumber\\
&\ge&\bigg[
-(\Sigma + \mu)|H_1|^2 - (\Sigma - \mu)|H_2|^2 + 2\xi\Sigma
\bigg]^\infty_{-\infty}.
\end{eqnarray}
The BPS solution saturates this inequality by satisfying 
the BPS equation:
\begin{eqnarray}
\frac{1}{e^2}\Sigma' 
&=& -\left(|H_1|^2 + |H_2|^2 - 2\xi\right),
\label{bps_finite1}\\
H_1' &=& - \left(\Sigma + \frac{\mu}{2}\right) H_1,
\label{bps_finite2}\\
H_2' &=& - \left(\Sigma - \frac{\mu}{2}\right) H_2.
\label{bps_finite3}
\end{eqnarray}

In the limit of infinite gauge coupling 
$e\rightarrow \infty$, Eq.~(\ref{bps_finite1}) 
yields a condition 
\begin{eqnarray}
|H_1|^2 + |H_2|^2 = 2\xi,
\end{eqnarray}
and Eq.~(\ref{bps_finite2}) and Eq.~(\ref{bps_finite3}) 
yields 
\begin{eqnarray}
(H_1 - H_2)' = - \frac{\mu}{2} \left( H_1 - H_2 \right),
\quad
\Sigma = - \frac{(H_1 + H_2)'}{H_1 + H_2}.
\end{eqnarray}
These can be easily solved. 
In order to show that 
these BPS equations are equivalent to the BPS 
equations (\ref{BPSeq:Theta}) and (\ref{BPSeq:R}) 
obtained previously, 
we just have to change variables 
\begin{eqnarray}
H_1 = \sqrt{2\xi} \cos\frac{\Theta}{2},\quad
H_2 = \sqrt{2\xi} \sin\frac{\Theta}{2}.
\end{eqnarray}
Then the above BPS equations with infinite gauge coupling 
are rewritten as 
\begin{eqnarray}
\Theta' = \mu\sin\Theta,\quad
\Sigma = - \frac{\mu}{2} \cos\Theta.
\end{eqnarray}
We obtain the same BPS solution as (\ref{eq:BPSsol-CP1}) 
for the infinite gauge coupling 
\begin{eqnarray}
\Theta 
=
\sin^{-1}\tanh\mu(y-y_0) + \frac{\pi}{2},
\qquad 
\Sigma 
=
\frac{\mu}{2} \tanh\mu(y-y_0), 
\label{theta_infinite}
\end{eqnarray}
\begin{eqnarray}
H_1 = \sqrt{2\xi}
\left(
\frac{{\rm e}^{-\mu y}}{{\rm e}^{\mu y} + {\rm e}^{-\mu y}}
\right)^{\frac{1}{2}},\quad
H_2 = \sqrt{2\xi}
\left(
\frac{{\rm e}^{\mu y}}{{\rm e}^{\mu y} + {\rm e}^{-\mu y}}
\right)^{\frac{1}{2}}.
\end{eqnarray}

Let us now turn to the case of finite gauge coupling. 
We have to solve the above three BPS equations 
Eq.(\ref{bps_finite1}), (\ref{bps_finite2}) and 
(\ref{bps_finite3}) for finite gauge coupling $e$.
Let us change variables 
\begin{eqnarray}
H_1 = \sqrt{2\xi}{\rm e}^{\rho}\cos\frac{\Theta}{2},
\quad
H_2 = \sqrt{2\xi}{\rm e}^{\rho}\sin\frac{\Theta}{2}.
\label{ansatz_finite}
\end{eqnarray}
Then the above BPS equations reduce to
\begin{eqnarray}
\frac{1}{e^2}\Sigma' 
= - 2\xi\left({\rm e}^{2\rho} - 1\right),\quad
\rho' = - \Sigma - \frac{\mu}{2}\cos\Theta,\quad
\Theta' = \mu\sin\Theta.
\end{eqnarray}
Notice that the equation for $\Theta$ is the same 
as the equation for the infinite 
gauge coupling. 
The solution has already given in Eq.(\ref{theta_infinite}).
Combining the first two equation with this solution 
(\ref{theta_infinite}), we obtain an equation
for $\rho$:
\begin{eqnarray}
\rho'' - \frac{2}{\alpha^2}\left({\rm e}^{2\rho} - 1\right) 
= \frac{1}{2\cosh^2(u-u_0)},
\end{eqnarray}
where $u \equiv \mu y,\ 
\alpha^2 \equiv\frac{\mu^2}{\xi e^2}$ 
and a prime denotes a derivative in terms of $u$.
The several exact solutions have already been found 
for several integer $\alpha$
\cite{Isozumi:2003rp}.
For example, in the case of $\alpha=2$ we obtain 
\begin{eqnarray}
\rho = \frac{1}{2}
\log\left(\frac{\cosh\mu(y-y_0)}{1+\cosh\mu(y-y_0)}\right).
\end{eqnarray}
Then we find as shown in Fig.5
\begin{eqnarray}
H_A^{(\alpha=2)} 
&=& \sqrt{\frac{\cosh\mu(y-y_0)}{1+\cosh\mu(y-y_0)}}\times
H_A^{(\alpha=0)},\\
\Sigma^{(\alpha=2)} &=& 
\Sigma^{(\alpha=0)} 
- \frac{\mu}{2}\frac{\tanh\mu(y-y_0)}{1+\cosh\mu(y-y_0)}.
\end{eqnarray}
\begin{figure}[t]
\begin{center}
\includegraphics[width=8cm]{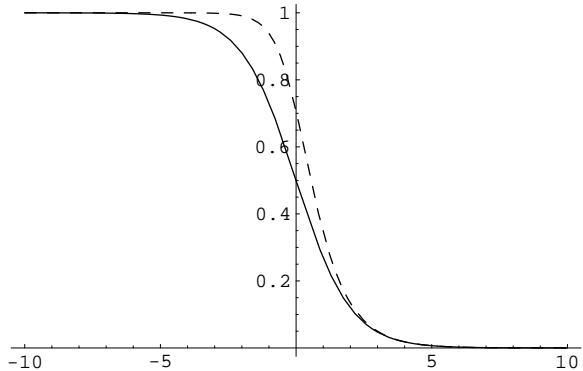}
\caption{The solid line is a BPS solution $H_1^{(\alpha=2)}$ 
with a finite gauge coupling $\alpha=2$ 
and the broken line is a BPS solution $H_1^{(\alpha=0)}$ 
with the infinite gauge coupling $\alpha=0$.}
\label{bps_finite}
\end{center}
\end{figure}


Let us finally explore non-BPS domain wall solutions in the 
case of finite gauge coupling. 
With the parametrization in Eq.(\ref{ansatz_finite}), 
the Lagrangian can be rewritten as 
\begin{eqnarray}
{\cal L} \!\!&=&\!\! - \frac{1}{2e^2}\Sigma'{^2} 
- 2 \xi {\rm e}^{2\rho}
\left(\rho'{^2} + \frac{\Theta'{^2}}{4}\right)
- V,\\
V \!\!&=&\!\! 2e^2\xi^2\left({\rm e}^{2\rho} - 1\right)^2 
+ 2\xi {\rm e}^{2\rho}
\left(\Sigma^2 + \frac{\mu^2}{4} 
+ \mu \Sigma \cos\Theta\right).
\end{eqnarray}
The equations of motion is given by
\begin{eqnarray}
&&\alpha^2 \tilde\Sigma'' 
- 2 {\rm e}^{2\rho}\left(2\tilde\Sigma 
+ \cos\Theta\right) = 0,\\
&&\rho'' + \rho'{^2} - \frac{\Theta'{^2}}{4} 
- \frac{2}{\alpha^2}\left({\rm e}^{2\rho}-1\right)
- \left(\tilde\Sigma^2 + \frac{1}{4}
+\tilde\Sigma \cos\Theta\right) = 0,\\
&&\frac{\Theta''}{2} + \rho'\Theta' 
+ \tilde\Sigma\sin\Theta = 0,
\end{eqnarray}
where we define $\tilde\Sigma = \Sigma/\mu$ and 
a prime denotes a derivative in terms of $u=\mu y$ 
in the rest of this subsection. 
One has to solve these coupled equations to obtain 
non-BPS solutions at finite gauge coupling. 
So far we have not succeeded to obtain a new solution. 
Let us finally examine the limit of infinite gauge 
coupling $\alpha \rightarrow 0$ of these equations. 
In such a limit the first and the second equations yield
\begin{eqnarray}
\tilde\Sigma = - \frac{1}{2}\cos\Theta,\quad
\rho = 0.
\end{eqnarray}
Inserting these into the last equation, we obtain 
the same equation as (\ref{eq:BPS-eq-theta}) 
\begin{eqnarray}
\Theta'' = \sin\Theta \cos\Theta.
\end{eqnarray}

\section{Stability}
\label{sc:stability}
\subsection{Stability under small fluctuation}
\label{sc:stability-small-fluc}
To examine the stability of the exact non-BPS domain wall 
solution (\ref{nbps_cp1}) in the massive 
$T^\star(\mathbb{CP}^1)$ 
nonlinear sigma model in five dimensions, 
we first study small fluctuations 
$(r,\theta,\varphi, \omega)$ 
around the background 
$\Theta_0(y) = 
{\rm am}\left(\frac{\Lambda}{k}(y-y_0),k\right) 
+ \frac{\pi}{2}$ and $\Phi_0 = {\rm const.}$ 
\begin{eqnarray}
\Theta(x^m,y) 
&\!\!\!=&\!\!\! \Theta_0(y) + \theta(x^m,y),
\quad
\Phi(x^m,y) = \Phi_0 + \varphi(x^m,y), 
\\
R(x^m,y)&\!\!\! =&\!\!\! r(x^m,y),
\quad 
\Omega(x^m,y) = \omega(x^m,y). 
\end{eqnarray}
The part of the Lagrangian quadratic in the fluctuations 
is decomposed into a sum 
for each fields 
\begin{equation}
{\cal L}^{(2)}_{\rm boson} 
=
{\cal L}^{(\theta,2)}_{\rm boson} + 
{\cal L}^{(\varphi,2)}_{\rm boson} + 
{\cal L}^{(r,2)}_{\rm boson} , 
\label{eq:quad-Lag}
\end{equation}
\begin{equation}
{\cal L}^{(\theta,2)}_{\rm boson}
=
 \int dy\ 
\xi\left\{- \frac{1}{2}\partial^M\theta\partial_M\theta 
- \frac{\mu^2}{2}\cos2\Theta_0 \theta^2
\right\}, 
\label{eq:quad-Lag-theta}
\end{equation}
\begin{equation}
{\cal L}^{(\varphi,2)}_{\rm boson} = 
\int dy\ 
\xi\left\{
- \frac{1}{2}\sin^2\Theta_0 
\partial^M\varphi\partial_M\varphi
\right\}, 
\label{eq:quad-Lag-phi}
\end{equation}
\begin{equation}
{\cal L}^{(r,2)}_{\rm boson} =
\int dy\ 
 \frac{1}{\xi}\left\{  
-\frac{1}{2}\partial^Mr \partial_Mr 
- \frac{1}{2}\left(\mu^2 + \frac{1}{2}\Theta_0'{^2} 
- \frac{\mu^2}{2}\sin^2\Theta_0\right)r^2
\right\}
.
\label{eq:quad-Lag-r}
\end{equation}
The linearized equations of motion read 
\begin{eqnarray}
&&
\left(\partial^m\partial_m + \frac{\partial^2}{\partial y^2} 
- \mu^2 \cos2\Theta_0 \right)\theta = 0,\\
&&\sin^2\Theta_0 \partial^m\partial_m\varphi 
+ \frac{\partial}{\partial y}
\left(\sin^2\Theta_0 \frac{\partial}{\partial y}\varphi\right) = 0,
\label{eq:varphi-linEOM}
\\
&&
\left(\partial^m\partial_m  + \frac{\partial^2}{\partial y^2}
-  \mu^2 - \frac{1}{2}\Theta_0'{^2} 
+ \frac{\mu^2}{2}\sin^2\Theta_0 \right) r = 0
\label{eq:r-linEOM}
.
\end{eqnarray}
Let us note that fluctuation of $\Omega$ disappears from 
the quadratic Lagrangian completely. 
Although this Lagrangian is sufficient to obtain 
light modes (those that become massless when radius 
goes to infinity), massive modes are expected from 
$\Omega$ if we wish to respect the four SUSY in the 
BPS limit of infinite radius. 
We describe an attempt to recover massive modes 
from the fluctuations of $\Omega$ by introducing 
a composite field $R\Omega$ in Appendix \ref{ap:omega-mode}. 

For $\theta$ and $r$, we can immediately 
define Shr\"odinger-type equations for 
mode functions $\psi_A^{(n)}$ with mass squared 
$m_{A,n}^2$ of effective 
fields on world volume as eigenvalues 
\begin{eqnarray}
\left[ - \frac{\partial^2}{\partial y^2} 
+ {\cal V}_A (y) \right]
\psi_A^{(n)}(y) 
=
 m_{A,n}^2 \psi_A^{(n)}(y),
\qquad 
A=\theta, 
r, 
\label{eq:mode-equation}
\end{eqnarray}
\begin{eqnarray}
{\cal V}_\theta(y) &=& 
\mu^2 \cos 2\Theta_0 = 
\mu^2\left\{ 2{\rm sn}^2\left(\frac{\mu}{k}(y-y_0),k\right) - 1\right\},
\label{spot_theta}
\\
{\cal V}_r(y) &=&
\mu^2 + \frac{1}{2}\Theta'_0{^2} - \frac{\mu^2}{2}\sin^2\Theta_0 
= \frac{1 + k^2}{2k^2} \mu^2
.
\label{spot_r}
\end{eqnarray}
For $\varphi$, we need to 
eliminate the first derivative 
of $\varphi$ in Eq.(\ref{eq:varphi-linEOM}) 
in order to obtain a Shr\"odinger-type equation. 
This is achieved by defining a field 
$\tilde \varphi$ 
\begin{eqnarray}
\tilde \varphi (x,y) = \varphi (x,y) \sin\Theta_0(y),
\end{eqnarray}
\begin{eqnarray}
\left(\partial^m\partial_m 
- \frac{\partial^2}{\partial y^2} 
- {\cos \Theta_0 \over \sin\Theta_0} \Theta_0{}''
+\left(\Theta_0{}'\right)^2
\right)\tilde\varphi
= 0. 
\end{eqnarray}
We now define mode functions 
$\psi_{\tilde \varphi}^{(n)}\equiv 
\psi_{\varphi}^{(n)} \sin\Theta_0$ for the 
potential ${\cal V}_{\varphi}(y)$ 
yielding mass squared $m_{\varphi,n}^2$ of 
the $n$-th effective fields 
 of $\varphi$ 
\begin{eqnarray}
\left[ - \frac{\partial^2}{\partial y^2} 
+ {\cal V}_\varphi (y) \right]
\psi_{\tilde \varphi}^{(n)}(y) 
=
 m_{\varphi,n}^2 \psi_{\tilde \varphi}^{(n)}(y),
\label{eq:phi-mode-equation}
\end{eqnarray}
\begin{eqnarray}
{\cal V}_{\varphi}(y) 
&=& {\cos \Theta_0 \over \sin\Theta_0}\Theta_0'' - (\Theta_0')^2
=
\mu^2
\left\{2 {\rm sn}^2\left(\frac{\mu}{k}(y-y_0),k\right) 
- \frac{1}{k^2} 
\right\}
.
\label{eq:pot-varphi}
\end{eqnarray}

We will first solve these eigenvalue equations 
(\ref{eq:mode-equation}) and (\ref{eq:phi-mode-equation}), 
and later study their normalizability 
in order to determine the physical 
modes among these solutions. 
If we replace $\mu$ by $\Lambda$, the Schr\"odinger potential 
(\ref{spot_theta}) is identical to 
the potential (\ref{seq_theta}) 
for $\theta$ in the sine-Gordon model in four dimensions. 
Therefore we obtain the same exact solutions for 
low-lying mode functions with normalization factors 
$N$'s : 
\begin{alignat}{2}
&m_{\theta,0}^2 = 0,\quad 
&\psi_\theta^{(0)} = N_\theta^{(0)}{\rm dn}
\left(\frac{\mu}{k}(y-y_0),k\right),
\label{eq:psi-massless}
\\
&m_{\theta,1}^2 = \frac{1-k^2}{k^2}\mu^2,\quad 
&\psi_\theta^{(1)} 
= N_\theta^{(1)}{\rm cn}\left(\frac{\mu}{k}(y-y_0),k\right),
\label{eq:psi-light}
\\
&m_{\theta,2}^2 = \frac{1}{k^2}\mu^2,\quad 
&\psi_\theta^{(2)} 
= N_\theta^{(2)}{\rm sn}\left(\frac{\mu}{k}(y-y_0),k\right)
.
\label{eq:psi-excited}
\end{alignat}
Eq.(\ref{eq:pot-varphi}) shows that the potential for $
\varphi$ is identical to 
that for $\theta$ except a constant shift : 
$
{\cal V}_{
\varphi} =
{\cal V}_\theta - \frac{1 - k^2}{k^2}\mu^2$. 
Therefore the same eigenfunctions as 
$\theta$ solve the eigenvalue problem for $
\varphi$ 
and the corresponding mass squared are 
shifted accordingly 
\begin{alignat}{2}
\label{eq:varphi-tachyon}
&m_{\varphi,-1}^2 = - \frac{1 - k^2}{k^2}\mu^2,
\quad 
&\quad \psi_{\tilde\varphi}^{(-1)} 
= N^{(-1)}_{\tilde\varphi}{\rm dn}
\left(\frac{\mu}{k}(y-y_0),k\right),
\\
\label{eq:varphi-massless}
&\quad m_{\varphi,0}^2 = 0,\quad 
&\psi_{\tilde\varphi}^{(0)} 
= N^{(0)}_{\tilde\varphi}{\rm cn}\left(\frac{\mu}{k}(y-y_0),k\right),\\
\label{eq:varphi-massive}
&\quad m_{\varphi,1}^2 = \mu^2,\quad 
&\psi_{\tilde\varphi}^{(1)} 
= N^{(1)}_{\tilde\varphi}{\rm sn}\left(\frac{\mu}{k}(y-y_0),k\right).
\end{alignat}
In contrast to the case of  $\theta$, 
the solution (\ref{eq:varphi-tachyon}) 
at first sight appears to indicate instability 
of the background solution for $k<1$ (with unit 
winding number), contrary to our expectations. 
However, we will see below that the possible 
tachyonic mode $\psi_{\tilde\varphi}^{(-1)}$ is 
not normalizable and unphysical in our case of 
$T^\star(\mathbb{CP}^1)$ model in five dimensions. 
The mode functions for $r$ can be completely obtained 
by plane waves with mass squared spectra 
\begin{equation}
 m_{r,n}^2 
= \frac{1 + k^2}{2k^2} \mu^2 
+ \left({n \over R}\right)^2
,\quad 
n =0, 1, 2, \cdots .
\label{eq:r-mass}
\end{equation}

Now let us determine physical modes by requiring 
the normalizability of these modes in the 
effective action. 
We expand 
the fluctuations 
in the quadratic Lagrangian (\ref{eq:quad-Lag}) 
by means of mode functions $\psi_A^{(n)}(y)$ with effective 
fields $f_A^{(n)}(x^m)$ 
as their coefficients ($A=\theta, \varphi, r
$) 
\begin{eqnarray}
\theta(x^m, y) &\!\!\!=&\!\!\! 
\sum_n \psi_\theta^{(n)}(y)f_\theta^{(n)}(x^m),\quad
\varphi(x^m, y) = \sum_n \psi_{
\varphi}^{(n)}(y) 
f_\varphi^{(n)}(x^m),
\label{eq:exp-theta-phi}
\\
r(x^m, y) &\!\!\!=&\!\!\! 
 \sum_n \psi_r^{(n)}(y) f_r^{(n)}(x^m),
\label{eq:exp-r}
\end{eqnarray}
where we use $\psi_{\varphi}^{(n)}(y)
= \psi_{\tilde \varphi}^{(n)}(y)/\sin\Theta_0(y)$. 
Apart from a trivial renaming of parameters, 
the quadratic Lagrangian (\ref{eq:quad-Lag-theta}) 
for $\theta$ is 
identical to that 
for $\theta$ in sine-Gordon model in four dimensions
\cite{Maru:2001gf}. 
Therefore we conclude that these modes of $\theta$ 
fluctuations are all physical. 
In the case of $k \le 1$ where we have quasi-periodic 
solution, 
we obtain no tachyonic mode, so that the
background configuration is stable for 
the fluctuation of $\theta$.
In the $k>1$ case, 
there is a tachyonic mode 
which destabilizes the background 
configuration. 
These results are identical to the four-dimensional case
\cite{Maru:2001gf}. 

The quadratic 
Lagrangian for $\varphi$ consists of kinetic 
and mass terms 
\begin{equation}
{\cal L}^{(\varphi,2)}_{\rm boson} 
=
{\cal L}^{(\varphi)}_{\rm kin} 
+ 
{\cal L}^{(\varphi)}_{\rm mass} 
\end{equation}
By means of the expansion (\ref{eq:exp-theta-phi}), 
the kinetic term is given by 
\begin{equation}
{\cal L}^{(\varphi)}_{\rm kin} 
=
-\frac{\xi}{2} \sum_{k,l} \int dy\ 
\partial^m f_\varphi^{(k)}(x) \partial_m f_\varphi^{(l)}(x)
\psi_{\tilde\varphi}^{(k)}(y) \psi_{\tilde\varphi}^{(l)}(y) 
.
\end{equation}
This gives the canonically normalized kinetic terms 
for effective fields 
\begin{equation}
{\cal L}^{(\varphi,2)}_{\rm boson} =-\frac{1}{2}
\sum_{n} \partial^m f_\varphi^{(n)}(x)\partial_m 
f_\varphi^{(n)}(x)
,
\end{equation}
if the following normalization condition 
is satisfied 
\begin{eqnarray}
\int dy\ \psi_{\tilde\varphi}^{(n)}(y) 
\psi_{\tilde\varphi}^{(l)}(y) 
= \frac{1}{\xi}\delta^{nl}.
\label{inner_phi}
\end{eqnarray}
The mass term, however, requires partial integrations 
in order to be transformed into 
the Schr\"odinger-type 
operator (\ref{eq:mode-equation}), 
\begin{eqnarray}
{\cal L}^{(\varphi)}_{\rm mass} 
&\!\!\!=&\!\!\!
- \frac{\xi}{2} 
\int dy\ 
\left(\sin\Theta_0
\partial_y\varphi \right)^2
=
- \frac{\xi}{2}
\int dy\ 
 \left(\partial_y\tilde\varphi 
-{({\rm sin}\Theta_0)' \over {\rm sin}\Theta_0}
\tilde\varphi\right)^2 
\nonumber\\
&\!\!\!=&\!\! 
- \frac{\xi}{2} 
\int dy\ \tilde \varphi 
\left(
-\partial_y^2 
+ {\cal V}_{\tilde\varphi}(y) \right)\tilde\varphi
- \frac{\xi}{2} 
\left[
\tilde\varphi {\partial \tilde\varphi \over \partial y} 
-{({\rm sin}\Theta_0)' \over {\rm sin}\Theta_0}
\tilde\varphi^2
\right]_{0
}^{2\pi L
}
.
\label{eq:phi-mass-norm}
\end{eqnarray}
The first term of the last equation reduces to the mass terms 
for effective fields by using the eigenvalue equation 
(\ref{eq:mode-equation}) and the orthonormality 
condition (\ref{inner_phi}) 
\begin{equation}
{\cal L}^{(\varphi)}_{\rm mass} 
=
- \frac{1}{2} 
\sum_{n}\left(m_{n}f_\varphi^{(n)}(x) \right)^{2}
.
\end{equation}
Therefore the normalizability of the mode functions 
is equivalent to the vanishing of 
the surface term in Eq.(\ref{eq:phi-mass-norm}). 
The solutions (\ref{eq:varphi-tachyon})--
(\ref{eq:varphi-massive}) 
for lower mass squared eigenvalues 
satisfy the orthonormality conditions. 
We find, however, that the surface term diverges 
for $\psi_{\varphi}^{(-1)}$ 
in Eq.(\ref{eq:varphi-tachyon}). 
Note that one has to evaluate the integral carefully 
by separating the integration 
region at $\mu(y-y_0)/k=K(k), 3K(k)$ in performing 
the partial integrations, since the integrand has singularities 
there. 
Therefore the possible tachyonic mode $\psi_{\varphi}^{(-1)}$ 
is not normalizable and hence unphysical. 
Other modes turn out to be normalizable and physical. 

Let us give a more intuitive explanation for the 
absence of the tachyonic modes. 
The contribution of the fluctuation $\varphi$ 
to the tension of the wall is nonnegative 
and vanishes if and only if 
$\varphi={\rm const.}$ 
\begin{eqnarray}
E_{\varphi} = \sin^2\Theta_0\left(\dot\varphi^2 
+ (\vec\nabla\varphi)^2\right) \ge 0.
\end{eqnarray}
Therefore any fluctuation $\varphi$ 
around our background configuration $\Phi_0=$constant 
increases the tension. 
This explains the classical stability of our 
non-BPS background configuration against the small 
fluctuation $\varphi$. 

Since the mode functions (\ref{eq:r-mass}) of the remaining 
fluctuations $r$ are obviously normalizable and 
consist of massive modes only, 
we conclude that our non-BPS two-wall solution is 
stable against small fluctuations. 

\subsection{Large fluctuations and topological aspect}
In this subsection we will show the instability of our 
non-BPS solution with respect to large fluctuations, 
by considering the topology of the model. 
Especially we would like to clarify differences between 
the four SUSY 
sine-Gordon model in four dimensions 
and the eight SUSY 
$T^\star(\mathbb{CP}^1)$ model in five dimensions. 
We will also propose to use a model with the 
double cover of $T^\star(\mathbb{CP}^1)$ 
manifold to assure 
the topological stability of our non-BPS solution. 

Let us first recall the situation of the 
four-SUSY sine-Gordon model.
In the sine-Gordon model, the real part $\Theta$ of 
the chiral scalar field is naturally a 
compact variable, taking values on 
$\Theta \in [0,2\pi) \simeq S^1$. 
Then the model acquires a topological quantum number 
$\pi_1(S^1)$ if we compactify the base 
space $y \sim y+2\pi L$ with the radius 
$2\pi L \equiv 4nkK(k)/\Lambda,  (n \in \mathbb{N})$. 
Since the non-BPS solution (\ref{nbps_sine}) is 
quasi-periodic for $k<1$, 
the nontrivial topological 
quantum number $n \in \pi_1(S^1)$ 
assures the stability of 
the non-BPS solution 
(\ref{nbps_sine}) for $k<1$ 
even under large fluctuations. 
On the other hand, the non-BPS solution is periodic 
for $k>1$ and has a vanishing topological 
quantum number corresponding to $n$ pairs of 
walls and anti-walls. 
They are unstable even under small fluctuations. 
The cases $k>1, k=1, k<1$ are illustrated and 
compared in Fig.\ref{nBPS_cp1}. 
The $k<1$ solution with $n \in \pi_1(S^1)$ 
has $n$ BPS walls and $n$ anti-BPS walls 
alternately. 
Since non-BPS configurations should have higher energy than 
the sum of two BPS walls, 
the BPS wall and the anti-BPS wall tend to exert repulsive 
force each other, resulting in wall positions at equal 
intervals on $S^1$. 
This intuitive explanation is in accord with our 
result that the non-BPS multi-wall configuration 
with $k<1$ is stable under large as well as small fluctuations.

Let us turn  our attention to the stability of 
the non-BPS solution (\ref{nbps_cp1}) 
in $T^\star(\mathbb{CP}^1)$ model. 
Although the field $\Theta$ in $T^\star(\mathbb{CP}^1)$ 
model parametrize a circle $S^1$, similarly to 
the $\Theta$ in the sine-Gordon model, 
this circle is just a great circle 
$S^2 \simeq \mathbb{CP}^1 \subset T^\star(\mathbb{CP}^1)$. 
Unlike $S^1$, $S^2$ is homotopically trivial 
 $\pi_1(S^2) = 0$.
Therefore the stability of 
our non-BPS solution (\ref{nbps_cp1}) is not supported 
by topological quantum numbers in the case of 
$T^\star(\mathbb{CP}^1)$ model.  
Although we have already verified that 
our non-BPS solution (\ref{nbps_cp1}) is stable against 
the small fluctuations, we still need to examine 
a possibility that the solution may 
be unstable under large fluctuations. 

To verify the instability under large deformations, 
we will examine a continuous deformation 
which makes the wall path shrinking to a point on $S^2$, 
in the spirit of variational approach. 
We will verify below that the energy of such a configuration 
shows local minimum around our non-BPS solution 
but eventually leading to true vacuum configuration 
without walls after passing over a maximum. 
This at least shows the existence of a continuous 
deformation of our non-BPS solution leading to 
no walls at all. 

\begin{figure}[t]
\begin{center}
\includegraphics[width=8cm]{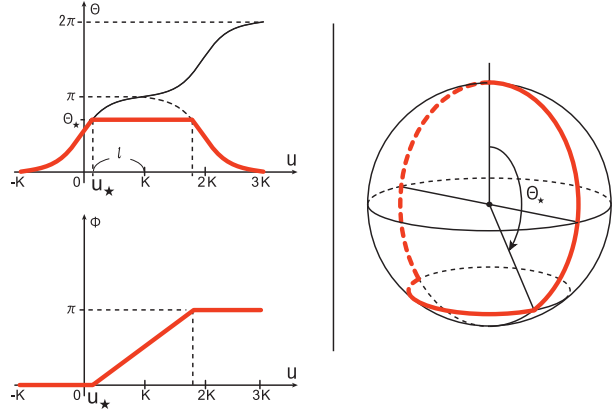}
\caption{The continuous deformation path is depicted.
In the left upper figure $\Theta$ is shown, in the left lower figure $\Phi$ is shown
and in the right the trial solution curve on the target space are depicted.}
\label{trial}
\end{center}
\end{figure}

For simplicity, we consider a path on $S^2$ which 
cuts off our non-BPS solution at $\Theta$, 
turns around a circle of $\Phi$ rotating by $\pi$ with 
the constant $\Theta$, and going to back through 
our non-BPS solution reflected at $\Theta=\pi$ : 
\begin{eqnarray}
\Theta_1(u;u_\star) &=& \left\{
\begin{array}{ll}
{\rm am}(u,k) + \dfrac{\pi}{2}&   -K < u \le u_\star,\\
\Theta_\star \equiv {\rm am}(u_\star,k) + \dfrac{\pi}{2}\quad
&  u_\star < u \le 2K - u_\star,\\
\dfrac{3\pi}{2} - {\rm am}(u,k)&  2K - u_\star < u \le 3K,
\end{array}
\right.\\
\Phi_1(u;u_\star) &=& \left\{
\begin{array}{ll}
0 &   -K < u \le u_\star,\\
\dfrac{\pi}{2}\dfrac{u-K}{K-u_\star} + \dfrac{\pi}{2}\qquad 
&  u_\star < u \le 2K - u_\star,\\
\pi &  2K - u_\star < u \le 3K,
\end{array}
\right.
\end{eqnarray}
where we have defined a variable 
\begin{equation}
u \equiv {\mu \over k}(y-y_0), 
\end{equation}
and $u_\star$ denotes the position in extra dimension 
corresponding to the value $\Theta=\Theta_\star$. 
This path is shown in Fig.\ref{trial}. 
This path $(\Theta_1,\Phi_1)$ 
interpolates the energy of our non-BPS 
solution $ (\Theta_0,\Phi_0=0)$ 
at $\Theta_\star \rightarrow \pi$ 
and the true vacuum $(\Theta=0,\Phi=0)$ 
without walls at 
 $\Theta_\star \rightarrow 0$. 
This is because the rotation of $\Phi$ by 
$\pi$ becomes irrelevant when 
$\Theta=0, \pi$, since all values of $\Phi$ corresponds 
to a single point at north or south poles. 

The energy of the configuration is given by
\begin{eqnarray}
{\cal E} = \int^{3K}_{-K} du\ \frac{\mu\xi}{2k}\left[
(\partial_u\Theta){}^2 
+ \left((\partial_u\Phi){}^2 + k^2\right)\sin^2\Theta\right].
\end{eqnarray}
\begin{figure}[t]
\begin{center}
\includegraphics[width=8cm]{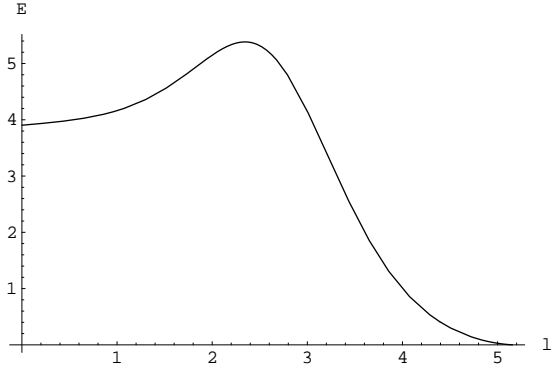}
\caption{The energy ${\cal E}$ of trial function as 
a function of $\ell\equiv K-u_\star$.}
\label{energy_trial}
\end{center}
\end{figure}
The energy of the above trial function is given by
\begin{eqnarray}
{\cal E}(u_\star) 
= 2{\cal E}_0(u_\star) + 2{\cal E}_1(u_\star),
\end{eqnarray}
where we define
\begin{eqnarray}
{\cal E}_0(u_\star) &\equiv&
\frac{\mu\xi}{2k} \int^{u_\star}_{-K}du\ 
\left[(\partial_u\Theta_0){}^2 + 
\left((\partial_u\Phi_0){}^2 + k^2\right)
\sin^2\Theta_0\right]\nonumber\\
&=& \frac{\mu\xi}{2k} \int^{u_\star}_{-K}du\ 
\left[2{\rm dn}^2(u,k) + k^2 - 1\right]\nonumber\\
&=& \frac{\mu\xi}{2k} \left[ E(u_\star) - E(-K) + (k^2 - 1)(u_\star + K)\right],
\end{eqnarray}
\begin{eqnarray}
{\cal E}_1(u_\star) &\equiv&
\frac{\mu\xi}{2k} \int^K_{u_\star} du\ 
\left[\Theta_1'{}^2 + \left(\Phi_1'{}^2 + k^2\right)
\sin^2\Theta_1\right]\nonumber\\
&=& \frac{\mu\xi}{2k} \int^K_{u_\star} du\ 
\left[ \left(\frac{\pi}{2}\frac{1}{K-u_\star}\right)^2 + k^2\right]
{\rm cn}^2(u_\star,k)\nonumber\\
&=& \frac{\mu\xi}{2k} \left[
\frac{\pi^2}{4} \frac{1}{K-u_\star} + k^2(K - u_\star) \right]
{\rm cn}^2(u_\star,k),
\end{eqnarray}
where $E(u)$ is the elliptic integral of the second kind.
To parametrize the path starting from our non-BPS solution, 
we introduce $\ell \equiv K - u_\star \in [0,2K]$ 
instead of $u_\star$. 
Then the total energy of our trial function in terms 
of $\ell$ is given by
\begin{eqnarray}
{\cal E}(\ell) = \frac{\mu\xi}{k}\left[
E(K\!-\ell)\! -\! E(-K) + (k^2 - 1)(2K-\ell)
+ \left(\frac{\pi^2}{4\ell} + k^2\ell\right)
{\rm cn}^2(K\!-\ell,k)
\right], 
\end{eqnarray}
which is shown in Fig.\ref{energy_trial}. 
We observe that the energy of the path of 
the continuous deformation has a local minimum at 
our non-BPS solution, in accordance with our result 
of no tachyon under small fluctuations. 
It then shows a maximum before reaching to the 
absolute minimum at the true vacuum. 
We regard this result as an evidence for the 
instability under large fluctuations. 

\begin{figure}[t]
\begin{center}
\includegraphics[width=8cm]{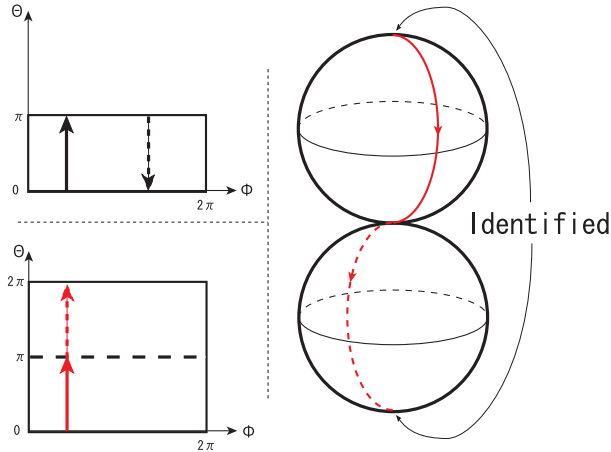}
\caption{Double cover of the base space of the $T^\star(\mathbb{CP}^2)$.}
\label{2s2}
\end{center}
\end{figure}
Although it may be enough to have metastability 
of our non-BPS solution with sufficiently long 
lifetime compared to the lifetime of our universe, 
it is certainly desirable to obtain topological 
stability under large fluctuations. 
We  can give the topological stability 
if we consider 
a double cover of the manifold $T^\star(\mathbb{CP}^1)$. 
For simplicity, let us explain in the case of the 
$\mathbb{CP}^1 \sim S^2$ 
model. 
If we take two spheres $S^2$, 
and identify the north (south) pole of one sphere 
 $S^2_{(1)}$ with the south (north) pole of 
the other sphere $S^2_{(2)}$, 
we obtain a double cover of $S^2$. 
This construction is illustrated in the right of Fig.\ref{2s2}.
The fundamental region and our non-BPS solution 
are shown in the left upper of Fig.\ref{2s2} for $\mathbb{CP}^1$ model and 
the left lower for the double cover of 
$\mathbb{CP}^1$. 
In this model, we can go over to the other sphere 
only through poles. 
Moreover, the path leading to the original sphere 
after going through the second sphere wind around the 
two spheres and cannot be deformed to nothing. 
Therefore we obtain the topological quantum number 
$\pi_1={\bf Z}$. 
On the other hand, all the local properties such as 
curvature are unaffected by such an identification. 
We believe that there is no global obstructions 
either. 
This consideration can be applied not only to the 
 $\mathbb{CP}^1$ model, but also to the 
 $T^\star(\mathbb{CP}^1)$ model. 
An interesting by-product of this double-cover model 
is that we can understand the reason why 
we have found no tachyons under small fluctuations 
of our non-BPS solution in the 
 $T^\star(\mathbb{CP}^1)$ model. 
This is because the quadratic Lagrangian of small fluctuations 
depends on only local properties of the model and should be 
identical to our stable model of double cover of 
$T^\star(\mathbb{CP}^1)$.

\section{SUSY breaking}
\label{sc:SUSYbreaking}

If we take a limit $L\rightarrow \infty$ 
with fixed $y_0$, we obtain a single BPS wall 
placed at $y_0$. 
On the other hand,if we take a limit $L\rightarrow \infty$ 
with fixed $y_0+\pi L$, we obtain a single anti-BPS wall 
placed at $y_0+\pi L$. 
The BPS wall solution preserves half of SUSY, say 
$Q_1$, and breaks the other half, $Q_2$. 
The anti-BPS wall solution preserves $Q_2$, and 
breaks $Q_1$. 
Therefore the coexistence of these two walls 
in our non-BPS two solution 
 breaks eight SUSY completely
\cite{Maru:2000sx}, \cite{Maru:2001gf}. 
From the brane world viewpoint, 
it is interesting to study how SUSY breaking 
effects are generated 
on a wall by the existence of the other walls. 
In usual SUSY breaking scenarios in the brane world, 
there are two 3-branes called the ``hidden brane" 
and the ``visible brane" 
in the higher dimensional spacetime . 
Once SUSY is broken by the vacuum expectation values 
of auxiliary fields of some supermultiplets 
in the hidden brane, SUSY breaking effects 
are mediated to the visible brane 
by bulk fields interacting with both branes. 
Then, soft SUSY breaking terms of MSSM 
fields on the visible brane 
are generated. 
In this framework, various fields have to 
be added on the hidden brane 
and/or in the bulk by hand to break SUSY and 
to transmit the SUSY breaking effects 
to our world. 
Furthermore, mechanisms of radius 
stabilization have to be specified 
to be phenomenologically viable. 
On the other hand, 
we have no need to add extra fields mentioned above 
since the non-BPS configuration itself breaks SUSY and 
the fields forming the non-BPS wall 
are responsible for SUSY breaking and 
its transmission to our world.  
As shown in the previous section, 
our non-BPS wall configuration is stable at least 
under small fluctuations. 
Therefore there is no need to introduce 
an additional mechanism 
to stabilize the radius. 
Moreover it aquires a topological stability 
under large fluctuations as well if we consider 
double cover of 
 $T^\star(\mathbb{CP}^1)$ model. 
In the light of these facts, 
it is worth while studying how SUSY breaking 
arises in our model.

Let us first understand the symmetry reason for 
the low-lying bosonic KK modes. 
In particular we are interested in those modes 
obtained in 
Eqs.(\ref{eq:psi-massless}), (\ref{eq:psi-light}), 
(\ref{eq:varphi-massless}), that are 
massless in the limit of large radius. 
Two zero modes $\theta^{(0)}$ and 
$\varphi^{(0)}$ are the Nambu-Goldstone modes corresponding 
to the broken 
spacetime and internal rotation symmetry, respectively.
On the other hand, $\theta^{(1)}$ represents fluctuations 
of relative distance of two walls\cite{Maru:2001gf}, 
which we call the breather mode. 
Similarly to the four-dimensional case, 
the wave function of  $\theta^{(0)}$ and 
 $\theta^{(1)}$ are peaked at two walls, 
but they have opposite sign around the anti-BPS wall 
located at $y=y_0+\pi L$. 
Thereforethe sum $(\theta^{(0)}+\theta^{(1)})/\sqrt2$ 
is localized at the BPS wall, and the difference 
$(\theta^{(0)}-\theta^{(1)})/\sqrt2$ at the anti-BPS wall. 
When we take the infinite radius limit of $k\rightarrow1$, 
the mass of breather mode vanishes. 

There are also fermionic zero modes which are 
dictated by symmetry reason. 
In the limit of $L\rightarrow \infty$ with 
$y_0+{\pi L \over 2}$ fixed, we obtain the BPS wall which breaks 
half of SUSY, $Q_2$. 
Therefore the corresponding Nambu-Goldstone fermion 
$f_0^{(2)}$ is localized on the BPS wall. 
Similarly, the anti-BPS wall is obtained by taking 
 the $L\rightarrow \infty$ limit with 
$y_0+{3\pi L \over 2}$ fixed, and the corresponding 
 Nambu-Goldstone fermion 
$f_0^{(1)}$ is localized on the anti-BPS wall. 
Since our non-BPS solution breaks 
all eight SUSY, we have two fermionic zero modes 
$f_0^{(1)}$ and $f_0^{(2)}$. 
The Nambu-Goldstone fermion $f_0^{(2)}$ is localized 
on the BPS wall, and another 
Nambu-Goldstone fermion $f_0^{(1)}$ is localized 
on the anti-BPS wall. 

If we fix $y_0+{\pi L \over 2}$ in taking the $L\rightarrow \infty$ 
limit, we find the mode 
 $(\theta^{(0)}+ \theta^{(1)})/\sqrt2$ 
as the surviving massless bosonic mode. 
This becomes the superpartner of the 
massless Nambu-Goldstone fermion $f_0^{(2)}$ 
corresponding to 
the SUSY broken by the BPS wall. 
On the other hand, if we fix $y_0+{3\pi L \over 2}$ in 
taking the $L\rightarrow \infty$ limit, 
the mode $(\theta^{(0)}-\theta^{(1)})/\sqrt2$ is the 
surviving massless boson which becomes the superpartner of 
the Nambu-Goldstone fermion $f_0^{(1)}$  
corresponding to the SUSY 
broken by the anti-BPS wall. 

These situations are precisely analogous to those of $\theta$ 
in the four SUSY 
sine-Gordon model in four dimensions
\cite{Maru:2001gf}. 
In that case, the BPS wall 
(anti-BPS wall) preserves two SUSY. 
Since the representation of two SUSY requires only a 
real scalar field, the real scalar field 
 $(\theta^{(0)}+ \theta^{(1)})/\sqrt2$ 
($(\theta^{(0)}- \theta^{(1)})/\sqrt2$) 
is sufficient as a scalar component of 
the low energy effective theory 
on the $1+2$ dimensional world volume. 
On the contrary, four SUSY should be preserved by 
the BPS wall in our five-dimensional model. 
The minimal representation of four SUSY requires 
a chiral scalar multiplet which contains 
a complex scalar field. 
Thus there must be one more real scalar mode in 
addition to $\theta^{(0)}$. 
This is supplied by $\varphi^{(0)}$, 
which is the massless Nambu-Goldstone boson 
corresponding to the broken 
internal rotation symmetry, since SUSY demands 
equal mass for the real and imaginary parts of 
the scalar component of the chiral multiplet. 
Another supporting evidence comes from the fact 
that the wave function $\tilde\varphi^{(0)}$ 
becomes identical to that of the wave function 
 $(\theta^{(0)}+ \theta^{(1)})/\sqrt2$ 
or  $(\theta^{(0)}- \theta^{(1)})/\sqrt2$ 
(apart from the sign), 
in the 
$L\rightarrow \infty$ limit 
with $y_0$ or $y_0+\pi L$ fixed. 
It is interesting to observe that the wave functions become 
identical to  $(\theta^{(0)}+ \theta^{(1)})/\sqrt2$ 
or  $(\theta^{(0)}- \theta^{(1)})/\sqrt2$, only if 
the redefined field $\tilde\varphi$ is used 
instead of the original field 
$\varphi$. 
It is not easy to recognize the localization 
in terms of the original field $\varphi$, 
since its wave function $\psi_{\varphi}^{(0)}$ 
is constant. 
However, we can easily see that the wave function 
 $\psi_{\tilde \varphi}^{(0)}$ in 
Eq.(\ref{eq:varphi-massless}) is identical to 
that for $\psi_{\theta}^{(0)}$ in 
Eq.(\ref{eq:psi-massless}) and is localized, 
because of 
the nontrivial weight for the inner product of
$\varphi$.

In the limit of the infinite distance of walls 
$L\rightarrow \infty$ ($k \to 1$), 
both fermionic and bosonic light modes 
become massless. 
However, as the walls approach each other, 
the bosonic field $\theta^{(1)}$ acquires 
a nonvanishing mass 
squared because of SUSY breaking. 
The mass splitting 
$\Delta m^2 \equiv m^2_{{\rm boson}} - m^2_{{\rm fermion}}$ 
is simply given by 
$\Delta m^2 = m^2_{\theta,1}=\frac{1-k^2}{k^2}\mu^2$. 
This mass splitting can be related to the distance between 
the walls 
by noting $2 \pi \mu L = 4kK(k)$ where $K(k)$ 
is the complete elliptic integral 
of first kind. 
In the limit $k \to 1$, we obtain 
$K(k) \to 
\frac{1}{2}{\rm log} \left( \frac{1}{1-k^2} \right)$, 
leading to
\begin{equation}
\Delta m^2 = \mu^2 \frac{e^{-\pi \mu L}}{1-e^{-\pi \mu L}} 
\simeq \mu^2 e^{-\pi \mu L}. 
\label{splitting}
\end{equation}
The mass splitting is exponentially suppressed 
as a function of 
the distance $\pi L$ between walls. 
If one considers the case with $L \to \infty$, 
the mass splitting vanishes, 
one recovers the single wall case 
which preserves the four SUSY. 
In this way, the result (\ref{splitting}) is consistent 
with our physical understanding. 
This result is also phenomenologically fascinating 
in that the low SUSY breaking scale can be 
naturally generated 
from the five-dimensional Planck scale 
$\mu \sim {\cal O}(M_5)$ 
without an extreme fine-tuning of parameters.

The qualitative features of SUSY breaking is 
the same for the sine-Gordon model 
in four dimensions \cite{Maru:2001gf}. 
The exponentially suppressed mass splitting 
has already been obtained 
in the sine-Gordon model, which seems to be 
generic in this kind of models. 
The difference is the multiplet structure on a wall. 
In the sine-Gordon model in four dimensions, 
the multiplet is real since the three-dimensional 
theory on a wall preserves only two SUSY. 
In the present $T^\star(\mathbb{CP}^1)$ model 
in five dimensions, 
the multiplet should contains a complex scalar 
since the  four-dimensional theory on a wall 
preserves four SUSY.

The exponentially suppressed mass splitting has 
also been discussed 
in a model of SUSY warped compactification \cite{GP}. 
The twisted boundary condition at $y=\pi L$ brane generates 
the tree level mass splitting of order $ e^{-\pi  L/l}/l$ 
where $l$ is a length scale of the $AdS_5$ and 
$L$ is a compactification radius. 
The suppression factor originates from the warp factor 
of the metric 
in the model \cite{GP} and the mass splitting vanishes 
in the flat limit 
$l \to \infty$. 
On the other hand, 
the suppression factor in our case comes from the 
nontrivial nature of 
the background configuration and 
the SUSY breaking effects are present already 
in the purely rigid SUSY theory.

We can embed our model with rigid SUSY into 
five-dimensional supergravity. 
In fact, we have considered a similar problem 
in the case of four-dimensional SUSY 
sine-Gordon model and found that 
the non-BPS two-wall solution is stable 
even when it is embedded into supergravity at least 
for weak gravitational coupling\cite{Eto:2003xq}. 
The massless Nambu-Goldstone modes are absorbed 
by a Higgs mechanism into massive gauge fields. 
The first massive boson (the breather 
mode) of the rigid SUSY model becomes the lightest 
scalar field, which is usually called radion, 
in the supergravity model. 
Since we found that massless modes are precisely those 
expected from the spontaneous breaking of global 
symmetries, and the remaining light fields are the breather 
modes corresponding to the fluctuations of the 
distance between the walls. 
This situation is completely analogous to the above 
four-dimensional sine-Gordon model. 
Therefore we anticipate that the same reasoning will be 
applicable for the stability of our solution 
embedded into supergravity : 
our non-BPS solution is stable even in the presence of gravity, 
and the breather mode gives the lightest scalar field, 
the radion if the gravitational coupling is weak. 
The only difference compared to four-dimensional model 
is that the field content is richer in order to represent 
the more symmetry, such as twice larger numbers of 
SUSY.

\section*{Acknowledgements}
We thank Muneto Nitta and Tetsuya Shiromizu for 
useful discussion. 
One of the authors (M.E.) gratefully acknowledges 
support from the Iwanami Fujukai Foundation and 
from a 21st Century COE Program at 
Tokyo Tech "Nanometer-Scale Quantum Physics" by the 
Ministry of Education, Culture, Sports, Science 
and Technology.
This work is supported in part by Grant-in-Aid for Scientific 
Research from the Ministry of Education, Culture, Sports, 
Science and Technology, Japan No.13640269 and 
for priority area ``Origin of Mass'' No.16028203 (NS) and 
by Special Postdoctoral Researchers Program at RIKEN (NM). 

\renewcommand{\thesubsection}
{\thesection.\arabic{subsection}}
\appendix

\section{Gamma matrix and spinor in 
$d=4,5$ 
dimensions}
\label{ap:gamma}
\subsection{Four dimensions 
}
\subsubsection{The Clifford algebra and the Lorentz algebra}
The Clifford algebra in four dimensions is of the form:
\begin{eqnarray}
\left\{\gamma_m,\gamma_n\right\} 
= - 2 \eta_{mn} \times \bm 1,
\end{eqnarray}
where $m,n$ run from 0 to 3 and 
$\eta_{mn} = {\rm diag.}(-1,1,1,1)$.
One of the representation of the Clifford algebra 
is given by
\begin{eqnarray}
\gamma_m = \left(
\begin{array}{cc}
0 & \sigma_m\\
\bar\sigma_m & 0
\end{array}
\right),
\label{eq:chiral-rep}
\end{eqnarray}
where $\sigma_m = (\bm 1,\sigma_i),\ \bar\sigma_m = (\bm 1,-\sigma_i)$.
The chirality matrix is defined as
\begin{eqnarray}
\gamma \equiv \gamma^0\gamma^1\gamma^2\gamma^3 
= \left(
\begin{array}{cc}
-i & 0 \\
0  & i
\end{array}
\right).
\end{eqnarray}
This chirality matrix anticommute with all the 
gamma matrices.

A representation of the Lorentz algebra is given by 
these gamma matrices:
\begin{eqnarray}
M_{mn} \equiv \frac{i}{4}\left[\gamma_m,\gamma_n\right].
\end{eqnarray}
These satisfy the Lorentz algebra
\begin{eqnarray}
\left[M_{kl},M_{mn}\right] = i 
\left(
\eta_{km} M_{ln} - \eta_{kn} M_{lm} 
- \eta_{lm} M_{kn} + \eta_{ln} M_{km}
\right).
\end{eqnarray}
Notice that if $\gamma_m$ satisfies the above Clifford 
algebra, $-\gamma_m,\gamma^*_m,-\gamma^*_m,
\gamma_m^\dagger,-\gamma_m^\dagger,\gamma_m^T,-\gamma_m^T$ 
also satisfy the same Clifford algebra.
These belong to the same equivalence class of the Clifford 
algebra, so there exist non singular 
matrices which intertwine $\gamma_m$ and others.
Let us define these intertwiners as
\begin{eqnarray}
\Df{A}\gamma_m\Df{A}{}^{-1} = \gamma_m^\dagger,\quad
\Df{B}\gamma_m\Df{B}{}^{-1} = - \gamma_m,\quad
\Df{C}{}^{-1}\gamma_m\Df{C} = - \gamma_m^T,\quad
\Df{D}{}^{-1}\gamma_m\Df{D} = - \gamma_m^*.
\end{eqnarray}
We can find other intertwiners by combining 
$\Df{A},\Df{B},\Df{C}$ and $\Df{D}$.
In our representation (\ref{eq:chiral-rep}) we obtain 
\begin{eqnarray}
\Df{A} = \gamma_0,\quad
\Df{B} = \gamma,\quad
\Df{C} = \left(
\begin{array}{cc}
i\bar\sigma_2 & 0\\
0 & i\sigma_2
\end{array}
\right),\quad
\Df{D} = \Df{C} \Df{A}{}^T
= \left(
\begin{array}{cc}
0 & i\bar\sigma_2\\
i\sigma_2 & 0
\end{array}
\right).
\end{eqnarray}
\subsubsection{Four component spinors: Dirac, Majorana, 
Weyl spinors}
Dirac spinor $\Psi$ is defined as a representation 
vector of the 
spinor representation 
of the Lorentz group:
\begin{eqnarray}
\Psi \rightarrow 
\exp\left(-\frac{i}{2}\theta^{mn}M_{mn}\right) \Psi.
\end{eqnarray}
Let us also define Dirac conjugate spinor of $\Psi$ by
\begin{eqnarray}
\bar\Psi \equiv \Psi^\dagger \Df{A} = \Psi^\dagger \gamma_0.
\end{eqnarray}
Because of the relation such as 
$\Df{A} M_{mn} \Df{A}{}^{-1} = M_{mn}^\dagger$, 
the Dirac conjugate spinor transforms as follows:
\begin{eqnarray}
\bar\Psi \rightarrow \bar\Psi 
\exp\left(\frac{i}{2}\theta^{mn}M_{mn}\right).
\end{eqnarray}
So we can easily find that $\bar\Psi\Psi$ is 
scalar under the Lorentz transformation.

The above representation $M_{mn}$ of the Lorentz 
algebra is not irreducible.
There are two ways to obtain irreducible representations 
from this representation. 
One of them is called Majorana spinor representation. 
To define the Majorana spinor,
we first need to define charge conjugate of the Dirac spinor by
\begin{eqnarray}
\Psi^C \equiv 
\Df{D}\Psi^* = \Df{C}\Df{A}{}^T\Psi^* = \Df{C}\bar\Psi^T.
\end{eqnarray}
Notice that the transformation law of the charge conjugate 
spinor $\Psi^C$ is the same as
$\Psi$ because of the relation such as 
$\Df{D}{}^{-1}M_{mn}\Df{D} = - M_{mn}^*$:
\begin{eqnarray}
\Psi^C \rightarrow 
\exp\left(-\frac{i}{2}\theta^{mn}M_{mn}\right) \Psi^C.
\end{eqnarray}
Taking this property into account, we can consistently 
define the Majorana spinor by
\begin{eqnarray}
\Psi = \Psi^C.
\end{eqnarray}
Another possibility of the irreducible 
representation of the Lorentz algebra is 
called Weyl spinor representation. 
In order to define the Weyl spinor, we first
introduce a projection operator:
\begin{eqnarray}
{\cal P}^{(\pm)} = \frac{1 \pm i\Df{B}}{2} 
= \frac{1 \pm i\gamma}{2}.
\end{eqnarray}
We can decompose the Dirac spinor into two kinds 
of Weyl spinors:
\begin{eqnarray}
\Psi^{(+)} \equiv {\cal P}^{(+)} \Psi,
\quad \Psi^{(-)} \equiv {\cal P}^{(-)} \Psi.
\end{eqnarray}
The representation matrix of the Lorentz group is given by
\begin{eqnarray}
M^{(\pm)}_{mn} \equiv {\cal P}^{(\pm)} M_{mn}.
\end{eqnarray}
Each of them also forms a representation of 
the Lorentz algebra. 
More explicitly we obtain 
\begin{eqnarray}
M^{(+)}_{mn} &=& 
\left(
\begin{array}{cc}
\frac{i}{4}\left(\sigma_m\bar\sigma_n 
- \sigma_n\bar\sigma_m\right) & 0\\
0 & 0
\end{array}
\right)
\equiv \left(
\begin{array}{cc}
\Sigma^{(+)}_{mn} & 0\\
0 & 0
\end{array}
\right),\\
M^{(-)}_{mn} &=& 
\left(
\begin{array}{cc}
0 & 0\\
0 & \frac{i}{4} \left(\bar\sigma_m\sigma_n 
- \bar\sigma_n\sigma_m\right)
\end{array}
\right)
\equiv \left(
\begin{array}{cc}
0 & 0\\
0 & \Sigma^{(-)}_{mn}
\end{array}
\right).
\end{eqnarray}
\subsubsection{Two component spinor: Weyl spinor}
In terms of the four component spinor notation, the Weyl spinors are represented as
\begin{eqnarray}
\Psi^{(+)} =
\left(
\begin{array}{c}
\psi_\uparrow\\
0\\
0
\end{array}
\right),\qquad
\Psi^{(-)} =
\left(
\begin{array}{c}
0\\
0\\
\psi_\downarrow
\end{array}
\right),
\end{eqnarray}
where both $\psi_\uparrow$ and $\psi_\downarrow$ are two component complex spinors.
It is obvious that the upper half components of $\Psi^{(+)}$
or the lower half components of $\Psi^{(-)}$ are enough for a representation of the Lorentz group.
Transformation law of these Weyl spinors is given by
\begin{eqnarray}
\psi_\uparrow \rightarrow \exp\left( - \frac{i}{2} \theta^{mn} \Sigma^{(+)} \right) \psi_\uparrow,\\
\psi_\downarrow \rightarrow \exp\left( - \frac{i}{2} \theta^{mn} \Sigma^{(-)} \right) \psi_\downarrow.
\end{eqnarray}
Notice that $\Sigma^{(+)\dagger}_{mn} = \Sigma^{(-)}_{mn}$.
So $\psi_\downarrow^\dagger\psi_\uparrow$ and 
$\psi_\uparrow^\dagger\psi_\downarrow$ are scalar under 
the Lorentz transformation.
Let us introduce a useful notation: dotted and undotted spinorial index.
First we define index of $\sigma_m$ and $\bar\sigma_m$ as follows:
\begin{eqnarray}
\sigma_m \equiv \sigma_{m\alpha\dot\alpha},\quad
\bar\sigma_m \equiv \bar\sigma_m^{\dot\alpha\alpha}.
\end{eqnarray}
This implies index structure of 
$\Sigma_{mn}^{(+)}$ and $\Sigma_{mn}^{(-)}$ as follows
\begin{eqnarray}
\left(\Sigma_{mn}^{(+)}\right)_\alpha{^\beta} 
&=& \frac{i}{4} \left(
\sigma_{m\alpha\dot\beta}\bar\sigma_n^{\dot\beta\beta} - 
\sigma_{n\alpha\dot\beta}\bar\sigma_m^{\dot\beta\beta}\right),
\\
\left(\Sigma_{mn}^{(-)}\right)^{\dot\alpha}{_{\dot\beta}} &=&
\frac{i}{4} \left(
\bar\sigma_m^{\dot\alpha\beta}\sigma_{n\beta\dot\beta}
- \bar\sigma_n^{\dot\alpha\beta}\sigma_{m\beta\dot\beta}
\right).
\end{eqnarray}
We use a convention where 
undotted spinor indices are contracted like $\searrow$, 
and dotted spinor indices like $\nearrow$. 
Thus we should take the index structure of 
$\psi_\uparrow$ and $\psi_\downarrow$ as 
\begin{eqnarray}
\psi_\uparrow = \psi_{\uparrow\alpha},\qquad
\psi_\downarrow = \psi_\downarrow^{\dot\alpha}.
\end{eqnarray}

The Lorentz algebra is isomorphic to $SL(2,C)$. 
The invariant tensor of $SL(2,C)$ is an anti-symmetric 
tensor 
$\epsilon^{\alpha\beta} = (i\sigma_2)^{\alpha\beta}$:
\begin{eqnarray}
\epsilon^{\alpha\beta}
\exp\left(-\frac{i}{2}\theta^{kl}
\Sigma_{kl}^{(+)}\right)_\alpha{^\gamma}
\exp\left(-\frac{i}{2}\theta^{mn}
\Sigma_{mn}^{(+)}\right)_\beta{^\delta} 
= \epsilon^{\gamma\delta}.
\end{eqnarray}
Therefore we find that
\begin{eqnarray}
\epsilon^{\alpha\beta}\chi_{\uparrow\beta}
\psi_{\uparrow\alpha} 
\equiv \chi_\uparrow^\alpha \psi_{\uparrow\alpha}
\end{eqnarray}
is a Lorentz scalar. 
Comparing $(\psi_\downarrow^{\dot\alpha})^\dagger
\psi_{\uparrow\alpha}$ and 
$\chi_\uparrow^\alpha\psi_{\uparrow\alpha}$, we can identify
\begin{eqnarray}
(\psi_\downarrow^{\dot\alpha})^\dagger 
= \psi_\uparrow^\alpha.
\end{eqnarray}

The Dirac spinor and Majorana spinor are 
represented by Weyl spinors as follows:
\begin{eqnarray}
\Psi^D = \left(
\begin{array}{c}
\psi_{\uparrow\alpha}\\
\chi_\downarrow^{\dot\alpha}
\end{array}
\right),\quad
\Psi^M = \left(
\begin{array}{c}
\psi_{\updownarrow\alpha}\\
\bar\psi_\updownarrow^{\dot\alpha}
\end{array}
\right),
\end{eqnarray}
where $\bar\psi_\uparrow$ is complex conjugate of 
$\psi_\uparrow$.
\subsection{Five dimensions}
\subsubsection{The Clifford algebra and the Lorentz algebra}
Let us begin with the Clifford algebra in five dimensions 
\begin{eqnarray}
\left\{\Gamma_M,\Gamma_N\right\} = - 2\eta_{MN} \times \bm 1\quad
\left(M,N = 0,1,2,3,4\right),
\end{eqnarray}
where $\eta_{MN} = {\rm diag.}\left(-1,1,1,1,1\right)$.
We can easily construct a representation of this algebra by 
using a representation of the 
Clifford algebra in four dimensions as follows:
\begin{eqnarray}
\Gamma_{M=m} = \gamma_m,\quad
\Gamma_{M=4} = \gamma.
\label{eq:five-d-gamma}
\end{eqnarray}
There is no chirality matrix in five-dimensional 
Clifford algebra, since the corresponding 
$\Gamma$ matrix is proportional to 
the unit matrix:
\begin{eqnarray}
\Gamma \equiv \Gamma^0\Gamma^1\Gamma^2\Gamma^3\Gamma^4 
= - \gamma^2 = \bm 1.
\end{eqnarray}
A representation of the Lorentz algebra is given by
\begin{eqnarray}
M_{MN} = \frac{i}{4}\left[\Gamma_M,\Gamma_N\right].
\end{eqnarray}

Notice that there is 2 equivalence classes in $d=5$ Clifford algebra, since
there are no regular matrices which intertwine $\Gamma_M$ and $-\Gamma_M$.
The components of such 2 equivalence classes are
\begin{eqnarray}
\left\{\Gamma_M,\Gamma_M^\dagger,\Gamma_M^T,\Gamma_M^*\right\},\quad
\left\{-\Gamma_M,-\Gamma_M^\dagger,-\Gamma_M^T,-\Gamma_M^*\right\}.
\end{eqnarray}
Let us define intertwiners as follows:
\begin{eqnarray}
\Ds{A}\Gamma_M\Ds{A}{^{-1}} = \Gamma_M^\dagger,\quad
\Ds{C}{^{-1}}\Gamma_M\Ds{C} = \Gamma_M^T,\quad
\Ds{D}{^{-1}}\Gamma_M\Ds{D} = \Gamma_M^*.
\end{eqnarray}
For our representation (\ref{eq:five-d-gamma}) with 
(\ref{eq:chiral-rep}) 
we find
\begin{eqnarray}
\Ds{A} = \Df{A},\quad
\Ds{C} = i\Df{B}{^{-1}}\Df{C},\quad
\Ds{D} = i\Df{B}{^{-1}}\Df{D}.
\label{eq:five-d-intertw}
\end{eqnarray}
\subsubsection{Dirac spinor}
Similarly to 
the four-dimensional case, the Dirac spinor 
can be defined as 
\begin{eqnarray}
\Psi \rightarrow \exp\left(- \frac{i}{2}
\theta^{MN}M_{MN}\right) \Psi,
\end{eqnarray}
where $\Psi$ has four complex components, namely 
eight real degrees of freedom. 
Dirac conjugate of $\Psi$ is also defined by
\begin{eqnarray}
\bar\Psi \equiv \Psi^\dagger \Ds{A}.
\end{eqnarray}
Because of $\Ds{A} = \Df{A}$ in 
Eq.~(\ref{eq:five-d-intertw}), 
we can easily verify the following transformation law
\begin{eqnarray}
\bar\Psi \rightarrow \bar\Psi \exp\left(\frac{i}{2}\theta^{MN}M_{MN}\right).
\end{eqnarray}
\subsubsection{Symplectic Majorana spinor}
Unlike the case of four dimensions, 
neither Majorana nor Weyl condition can be imposed
in the case of five dimensions. 
However, we can impose other condition, so-called symplectic 
Majorana condition.
Let us consider 2 Dirac spinor $\Psi^i\ (i=1,2)$ and write its Dirac conjugate as $\bar\Psi_i$.
We can define charge conjugate spinor by
\begin{eqnarray}
\Psi^C_i \equiv \Ds{C}\bar\Psi_i^T.
\end{eqnarray}
Using the property 
\begin{eqnarray}
\Ds{C}M_{MN}^T\Ds{C}{^{-1}} = - M_{MN},
\end{eqnarray}
we can easily verify that the transformation law of 
$\Psi^C$ is the same as that of $\Psi$.
So we can define the Symplectic Majorana spinor by
\begin{eqnarray}
\Psi^i = \varepsilon^{ij}\Psi^C_j 
= \varepsilon^{ij}\Ds{C}\bar\Psi_j^T.
\end{eqnarray}
In terms of the two component spinors, 
the symplectic Majorana spinors are 
expressed as
\begin{eqnarray}
\Psi^1 = \left(
\begin{array}{c}
\psi_{\uparrow\alpha}\\
\bar\psi_\downarrow^{\dot\alpha}
\end{array}
\right),\quad
\Psi^2 = \left(
\begin{array}{c}
\psi_{\downarrow\alpha}\\
-\bar\psi_\uparrow^{\dot\alpha}
\end{array}
\right).
\end{eqnarray}
The following relations can also be verified:
\begin{eqnarray}
\bar\Psi\Gamma^M\partial_M\Psi &=& \frac{1}{2}\bar\Psi_i\Gamma^M\partial_M\Psi^i,\\
\bar\Psi_i\Psi^i &=& 0,\\
\bar\Psi_i\Gamma_M\Psi^i &=&0.
\end{eqnarray}
In terms of the symplectic Majorana spinor, the 
internal $SU(2)$ symmetry is manifest.

\section{Massive modes of $\Omega$ }
\label{ap:omega-mode}
As discussed in subsect.\ref{sc:stability-small-fluc}, 
the field $\Omega$ disappears from the quadratic Lagrangian
if $(R, \Phi, \Theta, \Omega)$ are used 
as independent fields for fluctuations. 
To respect the preserved four SUSY, we 
should have even number of real scalar fields 
as fluctuations. 
Therefore we consider a composite field 
$\omega\equiv R\Omega$ 
as elementary excitations, instead of $\Omega$ 
itself. 
This may perhaps be related to the fact that 
$\Omega$ becomes meaningful only when $R\not=0$. 
In the limit of infinite radius $L\rightarrow \infty$, 
the non-BPS walls for the $T^\star(\mathbb{CP}^1)$ model 
should recover four SUSY. 
Therefore there should be four fields in the case of 
the $T^\star(\mathbb{CP}^1)$ model, which become 
degenerate at least in the limit of $L\rightarrow \infty$. 

Let us recall that 
the relevant part of Lagrangian ${\cal L}^{\Omega}$ 
for $\Omega$ gives a quadratic part 
${\cal L}^{(2, \omega)}$ for the redefined field 
$\omega\equiv R\Omega$ 
\begin{eqnarray}
{\cal L}^{\Omega} &=& -\frac{1}{2 \xi}
\left[
R^2 \partial_M \Omega \partial^M \Omega 
+2 R^2 {\rm cos} \Theta \partial_M \Phi \partial^M \Omega
\right] \\
\rightarrow 
{\cal L}^{(2, \omega)} 
&=& -\frac{1}{2\xi}
\left[
\partial_m \omega \partial^m \omega 
+ (-\frac{R_0'}{R_0}\omega + \omega')^2
\right],
\label{lag:omega}
\end{eqnarray}
where 
$R_0, \Phi_0$ denote the classical solutions 
of equations of motion for $R$ nd $\Phi$, and  
we used the fact that $\Phi_0$ is a constant. 
Let us note that our background solution may give 
nontrivial ${R_0'}/{R_0}$, since ${R_0}={R_0'}=0$. 
The linearized equation for $\omega$ is obtained 
from Eq.(\ref{lag:omega}) 
\begin{equation}
\left( \partial^m \partial_m 
+ \frac{\partial^2}{\partial y^2} 
- \left( \frac{R_0'}{R_0} \right)^2 
- \left( \frac{R_0'}{R_0} \right)'
\right) \omega = 0. 
\label{eq:omega}
\end{equation}
We will evaluate the potential term in Eq.(\ref{eq:omega}) 
by analyzing the equation of motion for $R$. 
Since we are interested in solutions close to $R_0\approx 0$, 
we can use the linearized equation of motion 
(\ref{eq:r-linEOM}) together with Eq.(\ref{spot_r}) 
for $R_0$ and multiply it by $R_0'$ to obtain 
by Eq.(\ref{eq:r-linEOM}) 
\begin{equation}
R_0' \partial^m \partial_m R_0 
+ \left( \frac{1}{2}(R_0')^2 \right)' 
-\frac{1+k^2}{2k^2} \mu^2 
\left( \frac{1}{2} R_0^2 \right)' = 0. 
\end{equation}
Since $R_0$ depends only on $y$, we obtain 
\begin{equation}
(R_0')^2 - \frac{1+k^2}{2k^2} \mu^2 R_0^2 = {\rm const}
\equiv A. 
\label{eq:constraint}
\end{equation}
Using this expression, it is easy to calculate the 
potential for $\omega$ in Eq.(\ref{eq:omega}). 
\begin{eqnarray}
\frac{R_0'}{R_0} 
&=& \pm \sqrt{\frac{1+k^2}{2k^2}\mu^2 + \frac{A}{R_0^2}}, \\
\label{ratio}
\left( \frac{R_0'}{R_0} \right)' &=& -\frac{A}{R_0^2}, 
\end{eqnarray}
leading to 
\begin{equation}
\left( \frac{R_0'}{R_0} \right)^2 
+ \left( \frac{R_0'}{R_0} \right)' 
= \frac{1+k^2}{2k^2} \mu^2. 
\label{pot-omega}
\end{equation}
By inserting the result (\ref{pot-omega}) into 
Eq.(\ref{eq:omega}) shows that 
the linearized equation of motion for 
$\omega\equiv R \Omega$ completely agrees with 
the  linearized equation of motion (\ref{eq:r-linEOM}) 
with (\ref{spot_r}) for $R$ in non-BPS case. 
Therefore, the mass spectra for $\omega$ and 
$R$ are identical. 
Note also that this result is valid irrespective 
of the value of the integration constant $A$. 

Noting that our solution is $R_0=0$, 
we have to check whether the solution obtained 
from Eq.(\ref{eq:constraint}) 
is consistent with $R_0=0$. 
It is easy to solve Eq.(\ref{eq:constraint}) as
\begin{equation}
R_0(y) = \sqrt{-\frac{2Ak^2}{(1+k^2)\mu^2}}{\rm cosh}
\left[ \pm \sqrt{\frac{1+k^2}{2k^2}\mu^2} 
(y + C) \right]~(C:{\rm const}). 
\end{equation}
When $A=0$, $R_0 = 0$ is reproduced. 
Therefore the choice of the background solution 
$R_0=R_0'=0$ corresponds to the integration constant 
$A=0$. 

Before closing Appendix B, we comment on BPS case. 
BPS case is obtained from the $k \to 1$ limit of non-BPS case. 
The agreement of the potential and the mass spectrum 
for $R$ and $\omega\equiv R\Omega$ in BPS case is obvious 
since the potential of $R$ and $R\Omega$ is 
$\frac{1+k^2}{2k^2}\mu^2 \to \mu^2~(k \to 1)$. 
On the other hand, 
the consistency of BPS solution with $R_0=0$ 
can also be seen from BPS equations. 
Recall that BPS equations for $\Theta, R$ have already 
been given by the first equation 
of Eqs.(\ref{BPSeq:Theta}) and (\ref{BPSeq:R}), 
\begin{equation}
\Theta_0' = \pm \mu {\rm sin}\Theta_0, \quad 
R_0' = \mp \mu R_0 {\rm cos}\Theta_0. 
\end{equation}
The BPS solution for $R_0$ is easily obtained 
by using the BPS solution (\ref{eq:BPSsol-CP1}) 
for $\Theta_0$ 
\begin{equation}
R_0(y;y_0) = C_0{\rm cosh}[\mu(y-y_0)]~(C_0:{\rm const}). 
\label{eq:BPSsol-R}
\end{equation}
Substituting this BPS solution into 
Eq.(\ref{eq:constraint}), 
the integration constant $A$ is fixed as 
\begin{equation}
A = R_0^2({\rm tanh}^2[\mu(y-y_0)] - 1) 
= \frac{-R_0^2}{{\rm cosh}^2[\mu(y-y_0)]} = -\mu^2 C_0^2. 
\end{equation}
To satisfy $R_0 = 0$, 
we find $C_0=0$ from (\ref{eq:BPSsol-R}), 
which again gives $A=0$. 
This result is consistent with Eq.(\ref{eq:constraint}).

Summarizing this appendix, 
we have shown that the mass spectra of $R$ and 
$\omega\equiv R\Omega$ agree completely. 
It is also shown that the BPS and non-BPS solutions of $R$ 
obtained in this appendix is consistent with our 
solution $R_0=0$.


\begin{thebibliography}{100}

  \bibitem{HoravaWitten}     
    P.~Horava and E.~Witten, 
     Nucl.\ Phys.\ {\bf B460}, 506 (1996) [hep-th/9510209]. 

 \bibitem{LED}N.~Arkani-Hamed, S.~Dimopoulos and G.~Dvali, 
             {\em Phys. Lett.} {\bf B429} (1998) 263 
             [hep-ph/9803315]; 
             I.~Antoniadis, N.~Arkani-Hamed, S.~Dimopoulos 
             and G.~Dvali, 
             {\em Phys. Lett.} {\bf B436} (1998) 257 
             [hep-ph/9804398]. 
 \bibitem{RandallSundrum}L.~Randall and R.~Sundrum, 
             {\em Phys. Rev. Lett.} 
             {\bf 83} (1999) 3370 [hep-ph/9905221]; 
             {\em Phys. Rev. Lett.} {\bf 83} (1999) 4690 
             [hep-th/9906064].

  \bibitem{DGSW}
    S.~Dimopoulos and H. Georgi, 
     Nucl.\ Phys.\ {\bf B193}, 150 (1981); 
    N.~Sakai, 
     Z.\ f.\ Phys.\ {\bf C11}, 153 (1981);
    E.~Witten, 
     Nucl.\ Phys.\ {\bf B188}, 513 (1981);
    S.~Dimopoulos, S.~Raby, and F.~Wilczek, 
     Phys.\ Rev.\ {\bf D24}, 1681 (1981).

\bibitem{cvetic1}
  M.~Cvetic, F.~Quevedo and S.~J.~Rey,
   Phys.\ Rev.\ Lett.\  {\bf 67}, 1836 (1991).
   
\bibitem{Bogomolny:1975de}
E.~B.~Bogomolny,
Sov.\ J.\ Nucl.\ Phys.\  {\bf 24}, 449 (1976)
[Yad.\ Fiz.\  {\bf 24}, 861 (1976)];
             M.~K.~Prasad and C.~H.~Sommerfield, 
             {\it Phys.\ Rev.\ Lett.\ } {\bf 35} (1975) 760.

 \bibitem{WittenOlive} E.~Witten and D.~Olive, 
             {\it Phys.\ Lett.\ } {\bf B78} (1978) 97.


\bibitem{Maru:2000sx}
N.~Maru, N.~Sakai, Y.~Sakamura and R.~Sugisaka,
Phys.\ Lett.\ B {\bf 496}, 98 (2000)
[arXiv:hep-th/0009023].

\bibitem{Maru:2001gf}
N.~Maru, N.~Sakai, Y.~Sakamura and R.~Sugisaka,
Nucl.\ Phys.\ B {\bf 616}, 47 (2001)
[arXiv:hep-th/0107204]; 
    N.~Maru, N.~Sakai, Y.~Sakamura, and R.~Sugisaka, 
     the Proceedings of the 10th Tohwa international symposium 
     on string theory, 
     American Institute of Physics, 607, pages 209-215, 
     (2002) [hep-th/0109087]; 
     ``SUSY Breaking by stable non-BPS configurations'', 
     to appear in the Proceedings of the Corfu Summer Institute
     on Elementary particle Physics, Corfu, September 2001 [hep-th/0112244].


 \bibitem{SS} N.~Sakai and R.~Sugisaka, 
              Phys.~Rev.~D65, (2002) 045010 [hep-th/0203142]. 

\bibitem{Alvarez-Gaume:1983ab}
L.~Alvarez-Gaume and D.~Z.~Freedman,
Commun.\ Math.\ Phys.\  {\bf 91}, 87 (1983).

\bibitem{SierraTownsend}
              G.~Sierra and P.K.~Townsend, 
               {\it Nucl.~Phys.} {\bf B233} (1984) 289. 

\bibitem{RT}
              M.~Ro\v{c}ek and P.~K.~Townsend, 
               {\it Phys.~Lett.} {\bf 96B} (1980) 72.
%
\bibitem{LR} 
              U.~Lindstr\"{o}m and M.~Ro\v{c}ek, 
               {\em Nucl.~Phys.} {\bf B222} (1983) 285;
              N.~J.~Hitchin, A.~Karlhede, U.~Lindstr\"{o}m and M.~Ro\v{c}ek, 
               {\it Comm.~Math.~Phys.} {\bf 108} (1987) 535. 

\bibitem{CF}  T.~L.~Curtright and D.~Z.~Freedman,
               {\it Phys.~Lett.} {\bf 90B} (1980) 71.

\bibitem{Gibbons:1987pk}
G.~W.~Gibbons, D.~Olivier, P.~J.~Ruback and G.~Valent,
Nucl.\ Phys.\ B {\bf 296}, 679 (1988).

\bibitem{Arai:2002xa}
M.~Arai, M.~Naganuma, M.~Nitta and N.~Sakai,
Nucl.\ Phys.\ B {\bf 652}, 35 (2003)
[arXiv:hep-th/0211103].

\bibitem{Arai:2003es}
M.~Arai, M.~Naganuma, M.~Nitta and N.~Sakai,
``BPS wall in N = 2 SUSY nonlinear sigma model with 
Eguchi-Hanson  manifold'', 
 ``Garden of Quanta''- 
  In honor of Hiroshi Ezawa, pages 299--325, 
World Scientific Pub. Co. Pte. Ltd. 
  Singapore, (2003), 
[hep-th/0302028].

\bibitem{GTT} J.~P.~Gauntlett, D.~Tong and P.~K.~Townsend,  
               {\it Phys.~Rev.} {\bf D64} (2001) 025010 
               [hep-th/0012178]. 

\bibitem{AFNS}
    M.~Arai,  S.~Fujita, M.~Naganuma, and N.~Sakai, 
  Phys.~Lett. {\bf B556} (2003) 192-202, 
  [hep-th/0212175]. 

\bibitem{EFNS}
    M.~Eto,  S.~Fujita, M.~Naganuma, and N.~Sakai, 
 Phys.Rev. {\bf D69} (2004) 025007,
   [hep-th/0306196]. 

\bibitem{Eto:2003xq}
M.~Eto, N.~Maru and N.~Sakai,
Nucl.\ Phys.\ B {\bf 673}, 98 (2003)
[arXiv:hep-th/0307206].

\bibitem{Eto:2003bn}
M.~Eto and N.~Sakai,
Phys.\ Rev.\ D {\bf 68}, 125001 (2003)
[arXiv:hep-th/0307276].

\bibitem{cvetic}
  M.~Cvetic, S.~Griffies and S.~J.~Rey,
   Nucl.\ Phys.\ B {\bf 381}, 301 (1992) [arXiv:hep-th/9201007]; 
  M.~Cvetic, S.~Griffies and H.~H.~Soleng,
   Phys.\ Rev.\ D {\bf 48}, 2613 (1993) [arXiv:gr-qc/9306005];
  M.~Cvetic and H.~H.~Soleng,
   Phys.\ Rept.\  {\bf 282}, 159 (1997) [arXiv:hep-th/9604090].


\bibitem{AraiNittaSakai}   M.~Arai, M.~Nitta, and N.~Sakai, 
 ``Vacua of Massive Hyper-K\"ahler Sigma Models 
of Non-Abelian Quotient'', 
 [hep-th/0307274].

\bibitem{KakimotoSakai}
 K.~Kakimoto and N.~Sakai, 
    ``Domain Wall Junction in ${\cal N}=2$ Supersymmetric 
    QED in four dimensions'', to appear in {\it Phys.~Rev.} 
    {\bf D68} (2003) 065005, [hep-th/0306077]. 

\bibitem{Isozumi:2003rp}
Y.~Isozumi, K.~Ohashi and N.~Sakai,
JHEP {\bf 0311}, 060 (2003)
[arXiv:hep-th/0310189].

\bibitem{GP}
T.~Gherghetta and A.~Pomarol,
Nucl.\ Phys.\ B {\bf 602}, 3 (2001) 
[arXiv:hep-ph/0012378].


\end{thebibliography}
\end{document}